\RequirePackage{fix-cm}
\documentclass{svjour3}
\smartqed

\DeclareUnicodeCharacter{2212}{-}

\usepackage{adjustbox}
\usepackage[utf8]{inputenc}
\usepackage{acronym}
\usepackage{graphicx}
\usepackage[T1]{fontenc}
\usepackage{epsfig}
\usepackage{txfonts}
\usepackage{amssymb}
\usepackage{siunitx}
\usepackage[backend=biber,sorting=none,minnames=3,defernumbers=true]{biblatex}
\usepackage{tocloft}
\usepackage{hyperref}
\usepackage{rotating}
\usepackage{pdflscape}
\usepackage{afterpage}

\journalname{ISSI Scientific Reports}

\begin{document}

\title{ESA Science Programme Missions: Contributions and Exploitation -- Herschel Observing Time Proposals}

\titlerunning{Herschel Observing Proposals}
\authorrunning{Pilbratt et al.}

\author{Göran L. Pilbratt \and Pedro Garc{\'i}a-Lario \and Arvind N. Parmar}

\institute{G. Pilbratt \at
Formerly at the Directorate of Science, ESA/ESTEC\\
The Netherlands\\
\email{glpil@pm.me}
\and
P. Garc{\'i}a-Lario \at
Directorate of Science, ESA, ESAC\\ 
Villanueva de la Ca\~nada, Madrid, Spain.\\
\and 
A. Parmar \at
Former Head of the Science Support Office\\
Directorate of Science, ESA, ESTEC\\
The Netherlands\\
\emph{Present address:}\\
Department of Space and Climate Physics\\
MSSL/UCL\\
Dorking\\ 
UK\\
}
                      
\date{Received:  date /  Accepted:  date}

\maketitle

\begin{abstract}
After an introduction to the ESA Herschel Space Observatory including a mission overview, science objectives, results and productivity we examine the process and outcomes of the announcements of observing opportunities (AOs). For Herschel, in common with other ESA observatories, there were no rules, quotas, or guidelines for the allocation of observing time based on the geographical location of the lead proposer's institute, gender, or seniority ("academic age"); scientific excellence was the most important single factor. We investigate whether and how success rates vary with these ("other") parameters. Due to the relatively short operational duration of Herschel -- compared to XMM-Newton and INTEGRAL -- in addition to the pre-launch AO in 2007 there was just two further AOs, in 2010 and 2011. In order to extend the time-frame we compare results with those from the ESA Infrared Space Observatory (ISO) whose time allocation took place approximately 15 years earlier.
\end{abstract}  

\section{\textbf{Introduction}}
\label{sec:Herschel_Introduction}

The Herschel Space Observatory was the \ac{ESA} far-infrared and sub-millimetre space observatory cornerstone facility \cite{2010Pilbratt}. It was equipped with a large passively cooled telescope and a science payload consisting of three complementary focal plane instruments located inside a superfluid helium cryostat. During an almost four year period in 2009--2013 it opened up and exploited the then poorly utilised spectral range $\sim$55--670\,$\mu$m for imaging and spectroscopy of a wide variety of astronomical targets spanning the solar system to distant galaxies, even peaking into the first billion years after the Big Bang. 

Herschel was the fourth and final cornerstone mission (CS4) of the Horizon 2000 Programme \cite{1984ESA}, following the already launched SOHO/Cluster (1995/2000) \cite{1995Fleck}, \cite{2001Escoubet}, XMM-Newton (1999) \cite{2001jansen}, and Rosetta (2004) \cite{2009Glassmeier} missions. The Herschel science payload was selected in 1997--99 and the actual construction of the spacecraft by a large industrial consortium commenced in 2001.

A self-contained and much more complete description of Herschel, including chapters on its origin, history, science objectives and results, technology and technical innovations, and management, can be found in \emph{Inventing a Space Mission: The Story of the Herschel Space Observatory} \cite{2017HerschelY}. A nicely illustrated summary can be found in the \emph{Herschel -- Science and Legacy} brochure \cite{2019HerschelSAL}.

\section{\textbf{Mission Overview}}
\label{sec:Herschel_MissionOverview}

Herschel was launched (together with the cosmic microwave background M3 mission Planck \cite{2010Tauber}) from \ac{CSG} Kourou by an Ariane V ECA rocket on 14 May 2009. The V188 dual launch was the first of its kind and was carried out flawlessly, releasing Herschel about 26 minutes after liftoff (and Planck a couple of minutes later) perfectly on track as planned. Exactly a month later, on 14 June 2009 while en-route towards its operational large amplitude quasi-halo orbit around the L2 Lagrangian point and still in the process of cooling down, the cryostat cover was successfully opened (and locked open) by manual commanding. For the first time Herschel could see the sky and performed a "sneak preview" of the galaxy M51 through its 3.5 metre diameter passively cooled Cassegrain telescope \cite{2009Doyle}, demonstrating good optical quality and correct alignment with focal plane units on the optical bench inside the cryostat. The telescope utilised (the then) novel silicon carbide technology and was the largest astronomical telescope ever operated in space until the launch of the \ac{JWST} on 25 December 2021.

The science payload was designed to provide photometric imaging and spectroscopic capabilities over the entire spectral range. This demanded complementary instruments employing a number of different enabling technologies. 

Two direct detection instruments each provided multi-band imaging photometry and low/medium spectral resolution imaging capabilities. The \ac{PACS} \cite{2010Poglitsch} covered the 55--210\,$\mu$m range, providing either dual-band imaging over a 1.75'\,x\,3.5' \ac{FOV} or 5\,x\,5 pixel spectroscopy with a spectral resolution of between 1500 and 4000, while the \ac{SPIRE} \cite{2010Griffin} covered the 194--672\,$\mu$m range, providing either triple-band imaging over a 4'\,x\,8' \ac{FOV} or imaging spectroscopy with 2.6' diameter \ac{FOV} and a spectral resolution of between 40 and 1000. Together \ac{PACS} and \ac{SPIRE} provided six wide photometry bands and low/medium spectroscopy covering the entire Herschel spectral range with a small overlap for cross-calibration.

The third instrument provided Herschel's very high spectral resolution capability. The \ac{HIFI} \cite{2010deGraauw} provided a single pixel on the sky with a bandwidth of up to 4 GHz anywhere between 480--1250\,GHz (625--240\,$\mu$m) or 1410--1910\,GHz (212--157\,$\mu$m) with a maximum spectral resolution beyond $10^6$.

For photometry the observer could choose either of \ac{PACS} or \ac{SPIRE}, or a special "parallel mode" of operating both of these simultaneously, useful for covering large areas on the sky by scanning. For spectroscopy one of the three instruments could be operated in any one of a number of modes offered. It turned out that over the mission lifetime very similar amounts of observing time were spent on photometry and spectroscopy.

Herschel was an observatory operated as a partnership between \ac{ESA}, the three instrument consortia and \ac{NASA}. ESA was responsible for all mission operations, conducted through the \ac{MOC} at \ac{ESOC} in Darmstadt, Germany. The \ac{MOC} was responsible for orbit maintenance, daily operations, uplink and downlink, including the health and safety of the spacecraft and instruments, supported as necessary by the other partners. For communication with the spacecraft either of ESA's New Norcia (Perth, Australia) or Cebreros (Avila, Spain) deep space antennas was used in a single scheduled 3-hour window every day. 

The Herschel \ac{SGS} \cite{2009Riedinger} was a partnership consisting of five elements: the \ac{HSC} located at the \ac{ESAC} near Madrid, Spain, an \ac{ICC} for each Herschel instrument provided by its consortium based at \ac{SRON}, Groningen, The Netherlands (HIFI), at \ac{MPE}, Garching, Germany (PACS), and at \ac{RAL}, Didcot, UK (SPIRE), and the \ac{NHSC} located at the \ac{IPAC} in Pasadena, California, USA. The \ac{HSC} performed scientific mission planning and was the prime interface between Herschel and the science community, supported by the partners. It provided information and user support related to the entire life-cycle of Herschel observations, including calls for observing time, the proposing procedure, proposal tracking, and supported the \ac{HOTAC} and the \ac{HUG} in executing their tasks. The \ac{HSC} also performed systematic pipeline data processing and hosted and populated the \ac{HSA} through which access to the Herschel data was, and still is, offered to the worldwide astronomical community, and provided the \ac{HIPE} for interactive data processing \cite{2010Ott}. The building of the necessary extensive software systems, in particular \ac{HIPE} \cite{2013Ott}, together comprising \ac{HCSS} was led by the \ac{HSC} and depended critically on contributions from all members of the \ac{SGS}.

The mission lifetime was limited by the continuously diminishing supply of the superfluid liquid helium cryogen which eventually was exhausted, with the last scientific observation being carried out on 29 April 2013. After some post-operations technical tests and eventually having been placed in its heliocentric "disposal orbit", the spacecraft was finally switched-off on 17 June 2013. Herschel has successfully conducted $\sim$37,000 science observations recommended by the \ac{HOTAC} in $\sim$23,400 hours of observing time, and additionally $\sim$6600 science calibration observations in another $\sim$2600 hours, all of which, in the form of a variety of derived data products at different levels of processing, are available through the \ac{HSA}.

\section{\textbf{Science Objectives and Results}}
\label{sec:Herschel_SciObjRes}

Given the considerable time between the 1984 selection of Herschel - called \ac{FIRST} at the time - and implementation naturally its science objectives were repeatedly subject to review and presented to the community for feedback, in the course of multiple studies undertaken. Specifically, this was the case ahead of the finalisation of its \ac{SMP} and the \ac{AO} for its science payload, both issued in the autumn of 1997, consisting of a special "hearing" with invited experts in September 1996 and the "Grenoble meeting" held in April 1997 \cite{1997Grenoble}.

Since Herschel was to "open up" a new (then) poorly observed part of the spectral regime, "the cool universe" (the blackbody peak corresponding to the spectral coverage is in the range 5--55 K), the idea was that the mission needed both to survey and follow-up on its own, while at the same time being lifetime constrained. Therefore the \ac{SMP} required that large observing programmes, called \ac{KPs}, should be selected and conducted early upfront in the mission. Special efforts were made to inform and engage the community early on, starting in earnest with the "Toledo conference" held in December 2000 \cite{2001Toledo}, in order to provide the community with ample time to prepare scientifically and organisationally to respond to the \ac{KP} \ac{AO}. This was also the meeting when Herschel got its name.

{\bf Main science objectives:} The top-level scientific areas foreseen for Herschel included wide-area galactic and extragalactic surveys, pertaining in particular to dust-enshrouded star formation throughout cosmic time; detailed studies of the \ac{ISM} in the Galaxy and other, primarily nearby resolvable, galaxies; observational astrochemistry of gas and dust in a variety of objects, as a tool for understanding physical and chemical processes throughout stellar lifecycles; and investigation of a wide variety of Solar System objects and their atmospheres. Herschel has delivered on all accounts, what follows is a -- most likely but unintentionally biased -- selection displaying the wide scope of scientific results.

{\bf Extragalactic surveys:} Two major photometric survey \ac{KPs} were conducted, supplemented by a number of other programmes. The number of "sub-millimetre" galaxies catalogued has been increased from some hundreds, or at most a couple of thousand, to in excess of half a million \cite{2016Valiante}, \cite{2018Maddox}, corresponding typically to 600+ sources per square degree on the sky. Herschel data suggest that although at all cosmic epochs the most vigorously star forming galaxies seem to be interacting galaxies \cite{2011Rodighiero}, \cite{2018Elbaz}, at least for redshifts up to z$\sim$2.5 (corresponding to a look back time of $\sim$11 Gyr) most of the star formation occurs in secularly star-forming galaxies where the infrared luminosity increases with redshift \cite{2011Elbaz}; sometimes referred to as the "galaxy main sequence", thus what a "normal" star-forming galaxy is depends on cosmic epoch. The generally greater productivity at earlier epochs appears simply to be associated with the availability of cold molecular gas - the raw material for making stars inferred from the thermal emission of dust that Herschel has observed - which was more plentiful at earlier epochs \cite{2010Dye}, \cite{2013Gruppioni}, \cite{2011Dunne}. As a population the galaxies detected in Herschel deep surveys resolve the cosmic infrared extragalactic background (which peaks at around 160 $\mu$m) into discrete sources \cite{2014Lutz}, removing any need for more imaginative contributions.

{\bf Early universe:} Herschel has also detected some galaxies in the first billion years (redshifts of z$\simeq$6 or greater) after the Big Bang \cite{2013Riechers}, and many more in the first couple of billion years \cite{2014Dowell}, \cite{2016Ivison}, \cite{2017Riechers}. These detections require massively star forming galaxies, with rates of thousands of solar masses per year (currently in the Galaxy the rate is about one solar mass per year). With the reservation that (a poorly known fraction of) these could be spatially unresolved groups of a few galaxies, the current understanding of structure formation and galaxy evolution in the very early universe is stretched.

{\bf \ac{ULIRGs}:} Spectral studies of relatively "local" \ac{ULIRGs}, essentially all of which are mergers of galaxies, have shown that the vast majority, if not all, of them display massive molecular outflows emanating away from them. These winds with velocities up to in excess of 1000 km s$^{-1}$ \cite{2010Fischer}, \cite{2014Gonzalez} are so powerful that they remove molecular material much faster than consumed by star formation. Unchecked they could remove all material to form any new stars in just millions or tens of million years, very short timescales on galaxy scales; an extreme form of "feedback" \cite{2011Sturm} indeed.

{\bf Supernova remnants:} A chance Herschel discovery was the remnant of a recent supernova, SN1987A, an unexpected by-product of a photometric survey of the Large Magellanic Cloud, a satellite galaxy of the Galaxy. The measured fluxes implied that enormous amounts of dust had been created in the explosion \cite{2011Matsuura}. If this amount of dust is formed by a typical supernova it would explain the amount of dust that Herschel has observed in the early universe. However, there is great uncertainty about how much dust actually survives from the "reverse shock" which is generated by the outgoing shock and travelling back towards the site of the explosion. This is a very interesting result, but the potential implications are not yet fully clear \cite{2015Matsuura}. Another serendipitous supernova related detection was that of the argon hydride ion $^{36}$ArH$^+$ through its J=1-0 and 2-1 lines in the Crab Nebula \cite{2013Barlow}. The $^{36}$Ar isotope is expected from the explosive nucleosynthesis in a core collapse supernova of a massive star, thus providing direct support for the Crab Nebula as a remnant of the supernova observed by the Chinese in 1054.

{\bf Low mass star formation:} Giant molecular clouds of interstellar matter near and far have been surveyed in multiple bands. Studies of low mass star formation in the nearby regions in the Gould Belt \cite{2010Andre} show that these display intricate networks of filamentary structures \cite{2010Menshchikov} with characteristic widths \cite{2011Arzoumanian}, \cite{2019Arzoumanian}. Herschel found them in all regions, some of which were forming stars \cite{2010Konyves}, \cite{2015Konyves}, \cite{2020Konyves}, but not all \cite{2010WardThompson}. A closer analysis \cite{2013Schneider} suggests that turbulence creates the structures, but that they need to be massive enough to become gravitationally unstable and fragment to form stars; Herschel data have enabled numbers to be asigned on what "massive enough" appears to be \cite{2013Konyves}, \cite{2015Andre} and have generated speculations about a new paradigm of (low mass) star formation \cite{2014Andre}. 

{\bf High mass star formation:} A complementary survey covered essentially all regions forming massive, OB, stars at distances less than 3 kpc from the sun, creating spectacular images \cite{2010Motte}, \cite{2010Zavagno}. A 1.2 THz wide \ac{HIFI} spectral survey in Orion, the closest massive star formation region, was conducted as part of a larger programme \cite{2010Bergin} and revealed a total of $\sim$13,000 lines from 79 isotopologues of 39 different molecules, with excellent agreement between observations and modelling \cite{2014Crockett} achieved. The chemical modelling \cite{2015Crockett} will represent a legacy for comparison with other chemical models and sources.

{\bf Galactic plane surveys:} In addition to specific galactic regions a photometric mapping of the entire 360$^{\circ}$ of the galactic plane with a latitude coverage of $\pm$1$^{\circ}$ has been conducted in about 1000 hours of observing time \cite{2010aMolinari}, \cite{2010bMolinari}; this is the largest chunk of Herschel observing time allocated to a single purpose. The survey takes the view of the Galaxy as a star formation engine \cite{2014Molinari}, attempts are made to understand star formation globally in the Galaxy, cataloguing several hundred thousand compact sources \cite{2016aMolinari}, \cite{2017Elia}, \cite{2021Elia}. Using this survey data the overall structure of the Galaxy \cite{2016bMolinari} and the large-scale properties of its interstellar medium have also been studied \cite{2018Elia}. Spectrally a major survey of the 158 $\mu$m ionised carbon, C$^{+}$, line along over 500 lines of sight through the galactic plane has been conducted \cite{2010Langer}. It has shown that there is significantly more galactic molecular hydrogen gas than currently inferred from carbon monoxide observations, the so-called "CO-dark" component \cite{2013Pineda}, \cite{2014Langer}. The C$^{+}$ line is also potentially a powerful star formation tracer \cite{2014Pineda}.

{\bf Infrared excess stars:} Herschel was not particularly suited to observe "normal" -- main sequence -- stars, like the sun, and was not normally used to, but an exception is the study of "infrared excess" which originates from dust surrounding a star (the "Vega phenomenon"). Two \ac{KPs} were conducted,  \cite{2010Eiroa}, \cite{2010Matthews}, together observing hundreds of nearby stars, finding and characterising infrared excess ("exo-Kuiper belts") to new levels of precision and resolution \cite{2010Sibthorpe}, \cite{2010Liseau}, \cite{2010Vandenbussche}, resulting in incidence levels for nearby FGK stars of $\simeq$ 21\% \cite{2013Eiroa}, \cite{2016Montesinos}. The nearby Fomalhaut displays a spectacular disc \cite{2012Acke} and in the young $\beta$ Pic the 69 $\mu$m band of crystalline olivine has been detected, exhibiting strikingly similar solid state properties to those of the dust emitted from the most primitive comets in the solar system \cite{2012deVries}.

{\bf Late type stars:} Stars in the process of formation accrete material while late-type ("dying") stars shed material, interacting with their surroundings in both cases. The morphology of the inner envelope and bow shock around the red supergiant Betelgeuse ($\alpha$ Ori) was imaged in all six Herschel bands \cite{2012aDecin}, displaying complicated multi-arch features in the bow shock region. A spectral survey was conducted on IRC+10216 (CW Leo). It showed a plethora of lines, many not previously observed, from many different molecules \cite{2010Decin}. Unexpectedly -- since for a "carbon-rich" star the oxygen was expected to be locked up mainly in CO and SiO -- no less than 39 ortho-H$_2$O and 22 para-H$_2$O lines, including low- and high-excitation lines from different parts of the envelope, were observed. Herschel has enabled important diagnostics of physical conditions and dynamics, for instance enabling to trace the mass loss of history of evolved stars, furthering the understanding of late stages of stellar evolution and the chemical enrichment of the interstellar medium \cite{2012bDecin}.

{\bf Water "trail":} Water is of particular interest, as a diagnostic line, but also as a coolant, and in its own right for its connection to life as we know it. Consisting of the two most common reactive elements (hydrogen and oxygen) in the universe, it forms on the surfaces of grains in molecular clouds, the birth places of stars. Thanks to its sensitivity, high spectral resolution, and capability to observe the water ground level transitions Herschel has managed to study the "water trail" from formation to planets \cite{2014vanDishoeck}. Herschel has detected water for the first time in a pre-stellar core in L1544 \cite{2012Caselli}, displaying an inverse P-Cyg profile characteristic of contraction. In star-forming regions water is a key tracer of dynamics and chemistry, with complicated line profiles indicative of both out- and inflows and shocks, but are typically dominated by broad features indicating bulk outflows \cite{2012Kristensen}. Herschel has provided the first observations of cold water vapour emission in the disc around the young star TW Hya \cite{2011Hogerheijde}, the vapour is in equilibrium with solid ice. The line was weaker than expected, but has enabled modelling of total water (vapour plus ice, where almost all the mass is in the ice) content. Later, the hydrogen deuteride (HD) fundamental (J=1-0) line was detected \cite{2013Bergin}, enabling a more direct modelling of the total disc mass, which is enough to form a planetary system like that of the solar system; despite having a relatively advanced, albeit still uncertain, age.

{\bf Water in the Solar System:} Water is an important constituent in comets, but has also been detected by the \ac{ISO} in 1997 in the stratospheres of all the four giant planets; a surprise discovery at the time the origin has been debated ever since. Herschel observations link the water observed in Jupiter's upper atmosphere to most likely originating from the 1994 impact of comet Shoemaker-Levy 9 \cite{2013Cavalie}, thus demonstrating water can be delivered by cometary impacts, and in Saturn's atmosphere originating from geysers on its moon Enceladus \cite{2011aHartogh}, \cite{2019Cavalie}. Furthermore, Herschel has made the first direct detection of water vapour around the dwarf planet Ceres \cite{2014Kuppers} in the main asteroid belt, and thus inside the "snowline".

{\bf Origin of water on Earth:} Comets were important Herschel targets. Early on water in the comet C/2008 Q3 Garradd \cite{2010Hartogh} was detected. For comets a particular interest was to measure the D/H-ratio of their water in order to compare with that of the oceans of the Earth, since the origin of the water on Earth is an open question subject to dispute. Before Herschel only Ivuna-type  carbonaceous (CI) chondrites had measured D/H ratios similar to that of Earth's water. However, Herschel measured the same ratio for the Jupiter family comet 103P/Hartley 2 (\cite{2011bHartogh}), which "made a splash" at the time. Then a higher ratio for the Oort cloud comet Garradd (in line with expectations for Oort cloud comets) was obtained \cite{2012BockeleeMorvan}, and yet later an upper limit, consistent with the Earth/Hartley 2 value, for the Jupiter family comet 45P/HMP \cite{2013Lis}. However, in 2016 Rosetta measured a very much higher ratio - higher even than for the Oort cloud comets - for its target, the Jupiter family comet 67P/CG \cite{2015Altwegg}. Not only is the issue of the provenance of Earth's water not resolved, but that of the D/H-ratio of comets has become more complicated than the pre-Herschel assumed simple picture.

{\bf \ac{TNO}s:} Herschel has also observed more than 130 of the approximately 1500 known \ac{TNO}s, the largest being Pluto and Eris, with >90\% detected for which photometric flux values have been obtained. These observations have transformed these cold "objects" outside the orbit of Neptune, to amazingly diverse "worlds" \cite{2012Vilenius}, \cite{2014Vilenius}, \cite{2014Lacerda}, \cite{2016Lellouch}. A few of the larger ones appear to have high albedos, indicative of "fresh" surfaces; could these be geologically active and able to resurface themselves, rather than being dead remnants of the formation of the solar system 4.5 billion years ago? The observations of Pluto made by NASA's New Horizons in 2015 may indicate this is true, at least for this particular \ac{TNO}.

{\bf ESA "web releases":} Most if not all of the above examples, and in fact many more, have been the subject of various ESA "web releases"; for a listing of all of these with links to stories and sources as appropriate in each case consult: \url{https://www.cosmos.esa.int/web/herschel/press-releases} and \url{https://www.cosmos.esa.int/web/herschel/latest-news}.

\section{\textbf{Science Productivity}}
\label{sec:Herschel_ScienceProd}

The wide scope of Herschel's science objectives and its spectral coverage made for a large user community, embracing the \ac{ISO} and Spitzer communities (the NASA mission Spitzer ran out of cryogen the day after Herschel was launched, a remarkable coincidence!) as well as radio astronomers including the \ac{SWAS} and Odin communities. This community made use of the data, produced science and published papers. Herschel also naturally has provided complementary views, to in particular radio, near infrared/optical and X-ray observations, some of them sometimes considered controversial from the perspectives of these other communities. Herschel has also complemented the Planck all-sky survey of the cosmic microwave background by providing higher angular resolution and spectral capabilities to "zoom in" on particular sources, doing astronomy and potentially aiding in the crucial removal the Planck "foregrounds".

Given the limited lifetime of the mission special efforts were made early during the in-flight mission to keep the community informed and updated of the actual demonstrated observatory capabilities, in particular in preparation for the first in-flight \ac{AO}, the AO1 (see below) that took place just within a year after the launch. Small numbers of observations from almost all \ac{KPs} were selected for early observing and release in what was called the \ac{SDP}. A number of meetings were organised, culminating in early May 2010 with the "Herschel First Results Symposium", immediately followed by pre-prints of accepted papers which were placed on the arXiv open-access archive. These were then published in two Special Issues of {\em Astronomy \& Astrophysics} dedicated to Herschel, and the scientific data were made publicly available to the entire community through the \ac{HSA}. 

\begin{figure*}[ht]
\centering
\includegraphics[width=1.0\textwidth,angle=0]{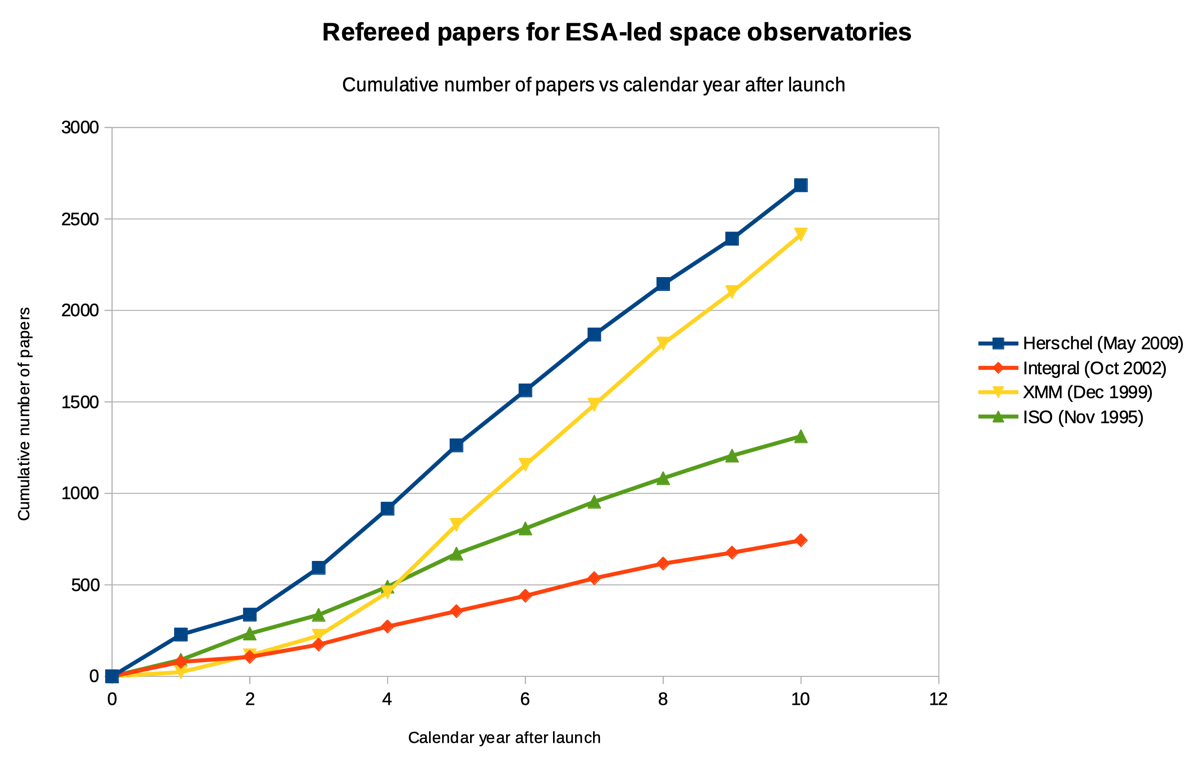}
\caption{Cumulative numbers of refereed papers for the four ESA-led space observatories ISO, XMM-Newton, INTEGRAL, and Herschel in the ten calendar years following their respective launch years (year 0 in the figure). ISO terminated observing in year 3 and Herschel in year 4 (in both cases due to the exhaustion of cryogen), while XMM-Newton and INTEGRAL are still observing more than 20 years after launch.}
\label{fig:Herschel_ESAledobs10yrpubs}    
\end{figure*}

This effort generated 152 papers for the {\em A\&A} Special Issues in 2010 alone, a good start that was further improved upon in the coming time and years. In fact, in the 10 calendar years following the launch year, Herschel has the highest number of publications of any of the ESA-led observatories to date, cf. Fig.~\ref{fig:Herschel_ESAledobs10yrpubs}. This is the more noteworthy given that by less than 4 years after launch Herschel was no longer observing (as was true for ISO about 2.5 years after its launch) while both XMM-Newton and INTEGRAL are still active space observatories more than 20 years after their respective launches. Interestingly - but outside the time range of the figure - the launch of Herschel appears to have stimulated the use of ISO data as the rate of ISO papers published increased at the time. 

Up to the end of 2021, there have been almost 3300 refereed Herschel publications  that make direct use of data from the mission including from its primary catalogues, or make quantitative predictions of results from the mission or describe Herschel, its instruments, operations, software or calibrations. In addition to a healthy publication record Herschel has also consistently scored well with respect to various other related parameters with a high number of citations, a considerable number papers with more than 100 citations and a high h-index (see \cite{2024pubs}). Herschel also has been scoring well in the ESA defined \ac{KPI}s regularly reported to the \ac{SPC}, and has generated a large number of \ac{PhD} theses over the years, mainly related to its science but also with a fraction related to its technical development. 

Given in particular the \ac{KPs} some Herschel programmes, or combinations of programmes, produced large data sets that were considered to potentially be of great legacy value. This generated a desire to produce suitable legacy datasets to made publicly available for the entire community, enhancing the value of the Herschel data and enable additional science to be performed. This would sometimes involve additional datasets and custom data processing and tools, requiring resources not available within the Herschel consortia on their own.

The European Union 7th Framework Programme (FP7) offered a potential solution by making funds available through competitive applications. Herschel teams were extremely successful, the two proposals HELP - "Herschel Extragalactic Legacy Project" \cite{2021Shirley} and VIALACTEA - "The Milky Way as a Star Formation Engine" \cite{2019Sciacca} each obtained the full score and were each awarded with 2.5 M\texteuro. In addition, DustPedia - "A Definitive Study of Cosmic Dust in the Local Universe" \cite{2017Davies} was awarded another 2 M\texteuro ~from FP7, and later SBNAF - "Small Bodies Near and Far" \cite{2018Muller} was awarded 1.5 M\texteuro ~from the follow-up EU Horizon 2020 programme. It is, however, unclear just exactly to what extent these projects have been fully successful in achieving their very ambitious goals.

\section{\textbf{Observing Time}}
\label{sec:Herschel_ObservingTime}

Herschel was designed to provide just under 20,000 hours for allocation to conduct scientific observations, out of which 32\% was \ac{GT}, mainly for the instrument consortia reflecting the requirements put on them to contribute to the science operations throughout the mission lifetime. The remainder was \ac{OT} which was open to the worldwide general astronomical community through a standard competitive \ac{AO} observing proposal submission process.

The \ac{KPs} were to be selected through a dedicated initial \ac{AO} before the launch, by necessity since these would be the first observing programmes to be conducted. This \ac{AO} thus would have to cope with the additional difficulty of an imprecise knowledge of the scientific capabilities of the observatory. To scientifically "validate" the selected \ac{KPs} given the actual performance before commencing full-scale observing was one of the reasons for the introduction of the \ac{SDP}. Additional \ac{AO}s would follow later in-flight. 

The first Herschel \ac{AO}, the dedicated "KP only AO", for proposals requiring substantial amounts ($>$100~hours) of observing time, was released in February 2007. Like all Herschel AOs it consisted of two parts; first a phase for \ac{KPGT}, followed by a phase for \ac{KPOT} after the accepted \ac{GT} programmes and associated observations had been identified and communicated. \ac{KPOT} proposals could only be submitted after the conclusion of the \ac{KPGT} phase. The entire KP \ac{AO} process was completed in February 2008, with the publication of the successful \ac{KPOT} programmes and observations. By coincidence it resulted in 21 \ac{KPGT} and 21 \ac{KPOT} proposals being awarded observing time. The first in-flight announcement (AO1) took place in 2010 with the initial guaranteed and subsequent open time phases being labelled GT1 and OT1. In 2011, the second, and final, announcement (AO2) was released with GT2 and OT2 phases as before; see Table~\ref{tab:Herschel_AOoverview} for an overview.

For the in-flight \ac{AO} OT phases the concept of priority 1 and priority 2 observing time was introduced. The background was that it had become clear that at all times it was necessary to have a pool of (at least) 6 months of observations in order to enable efficient scheduling. Ultimately this would mean that when Herschel ran out of helium there would be 6 months of accepted observations that would not be performed. A connected and important problem here, in particular regarding OT2, was a large uncertainty in the predicted lifetime. To cut a complicated story short, in the end approximately half of the priority 2 observations were actually executed. These were selected based primarily on grading (essentially all of the top third graded priority 2 proposals were observed, about half of the middle tier, and only a small fraction of the lower tier) but necessarily also on various additional constraints, in particular with respect to sky visibility which changed throughout the year.

In addition, for completeness, but for the purpose of this study negligible amounts of observing time were awarded as \ac{DDT}, "Must-Do", and "Filler" programmes. 

\begin{table}
\centering
\caption{Herschel AO overview showing dates, number of proposals, and awarded observing time. For completeness here both GT and OT are shown. The meaning of priorities p1 and p2 are described in the text. For the KP and GT AOs the concept of priorities was not applicable, the awarded time is equivalent to p1.}
\begin{tabular}{lccccccccc}
\hline \noalign{\smallskip}
\vspace{3pt}
AO & \multicolumn{2}{c}{Date of} & \multicolumn{4}{c}{Number of proposals} & \multicolumn{3}{c}{Time (hour)} \\
Phase & Issue & Deadline &  Sub-    &\multicolumn{3}{c}{Accepted} & \multicolumn{3}{c}{Awarded} \\
      &       &          & mitted & Total &  p1 &  p2 & Total & p1 & p2 \\
\hline \noalign{\smallskip}
KPGT & Feb 2007 & Apr 2007 &  21 &  21 &  21 & ... & 5879 & 5879 & ...  \\
KPOT & Feb 2007 & Oct 2007 &  62 &  21 &  21 & ... & 5379 &  5379 & ...   \\
GT1  & Feb 2010 & Mar 2010  & 33 &  33 &  33 & ... &  555 & 555 & ...  \\
OT1  & May 2010 & Jul 2010 & 577 & 241 & 176 &  65  & 6577 & 4987 & 1590  \\
GT2  & Apr 2011 & May 2011 &  32 &  32 &  32 & ... &  362 & 362  & ...  \\
OT2  & Jun 2011 & Sep 2011 & 531 & 373 & 181 & 192  & 7795 & 3420 & 4170  \\
\hline \noalign{\smallskip}
\end{tabular}
\label{tab:Herschel_AOoverview}
\end{table}
\subsection{\textbf{Herschel Observing Time Allocation Committee (HOTAC)}}
\label{sec:Herschel_HOTAC}

The \ac{HOTAC} was an external committee that evaluated proposals and made recommendations on the allocation of observing time to \ac{D-SCI}. Since the operational lifetime of Herschel crucially depended on its cryogen supply it had a limited, and for a major observatory relatively short, lifetime. The \ac{HOTAC} was formed with the hope that it could serve for the entire mission with a minimum of changes. It was decided that the \ac{GT} proposals should also be reviewed by the \ac{HOTAC}, for two reasons -- all proposals should represent science worthy of the limited availability of precious Herschel observing time, and by reviewing the \ac{GT} proposals the \ac{HOTAC} would obtain a good overview of, and familiarity with, the already allocated observing time in the various science areas, which was deemed useful when reviewing the \ac{OT} proposals. 

For operational reasons, and given the two (GT and OT) phases of each cycle, it was concluded that only one AO cycle per year could be supported, both from the \ac{HOTAC} as well as from the conducting the \ac{AO}, importantly including providing the necessary support to the \ac{HOTAC}, activities. For the OT phases the task of the \ac{HOTAC} was to grade the submitted proposals taking into account the scientific case, justification, merit and relevance of the proposed observation(s), and the potential contribution of the overall scientific return of the mission. The \ac{HOTAC} then made recommendations on the allocation of observing time, which could include modifications to what the proposer requested. The final decision concerning observing time allocation was performed by the \ac{D-SCI}.

The \ac{HOTAC} consisted of an overall chair and initially two parallel panels covering each of four science areas \ac{ISM/SF/SS}, \ac{S/SE}, \ac{G/AGN} and Cosmology. Each panel worked independently, and then a meeting at what became to be referred to as the "HOTAC Level", consisting of the overall chair and the panel chairs and vice-chairs, would derive the final overall \ac{HOTAC} recommendation. The \ac{HOTAC} chair would then communicate this to the \ac{D-SCI}. Since in the KP AO only a small number of proposals were expected, but collectively they would encompass a large amount of observing time, it was decided that for this AO all the work with all the proposals would be carried out at the \ac{HOTAC} level, consisting of 13 people, thus panel members were not recruited at this point.

For the AO1 and AO2 cycles, based on the KP AO proposal response in the various science areas, the \ac{HOTAC} structure was enlarged with a third \ac{ISM/SF/SS} panel. The \ac{ISM/SF/SS} and \ac{G/AGN} panels each had five members, and each of these two science areas supplied four people to the \ac{HOTAC} Level, while \ac{S/SE} and Cosmology panels each had four members, and each of these science areas supplied two people to the \ac{HOTAC} Level; these numbers approximately mirroring the proposal pressure. Thus, the Full \ac{HOTAC} consisted of 42 people, the \ac{HOTAC} Level of 13 people, in both cases including the overall Chair, see Table~\ref{tab:Herschel_HOTAC}. Although the objective was to keep as many people as possible from each \ac{AO} cycle to the next, each time approximately 15\% of the members were replaced, however, the \ac{HOTAC} Chair remained, and magically so did the gender distribution.

\begin{table}
\centering
\caption{A summary of the HOTAC membership. The HOTAC Chair is included in both the Full HOTAC and the HOTAC Level. The HOTAC Level members are included in the Full HOTAC. Although numbers appear persistent, actual compositions were not, with the exception of the Chair who was the same individual throughout (see text). \%f is the female percentage.}
\begin{tabular}{lcccccccccc}
\hline \noalign{\smallskip}
\vspace{3pt}
AO & Year & No. of & \multicolumn{3}{c}{Full HOTAC} & \multicolumn{3}{c}{HOTAC Level} & \multicolumn{2}{c}{HOTAC Chair} \\
 & Issued & Proposals  & Male & Female & \%f & Male & Female & \%f & ~~~Male & Female \\
\hline \noalign{\smallskip}
KPOT & 2007 & ~~62 &   - &   - &  - & 9 & 4 & 31 & ~~~0 & 1 \\
OT1  & 2010 &  577 &  29 &  13 & 31 & 9 & 4 & 31 & ~~~0 & 1 \\
OT2  & 2011 &  531 &  29 &  13 & 31 & 9 & 4 & 31 & ~~~0 & 1 \\
\hline \noalign{\smallskip}
\end{tabular}
\label{tab:Herschel_HOTAC}
\end{table}

\section{\textbf{Herschel Users' Group (HUG)}}
\label{sec:Herschel_HUG}

The \ac{HUG} had in common with the \ac{HOTAC} that it was an independent advisory group that was self-managed and was not a part of the Herschel \ac{SGS} (which had its own internal advisory groups). It was set up on the initiative of the Herschel Science Team (it was actually not required by, or even mentioned in, the \ac{SMP}) in the year of the launch, to provide a forum for (potential) general users of Herschel to provide input to the operations of the observatory as an astronomical facility on various matters affecting its scientific productivity and user friendliness. 

The \ac{HUG} conducted a preparatory teleconference in addition to a total of ten meetings in the period 2010--2016. It interacted directly with the Herschel user community, e.g., through polls and their own networks, it was supported by the \ac{PS} and the \ac{SGS}. It reported to the \ac{PS} who posted all the presentations made in the meetings and the minutes with the HUG advice on a dedicated \ac{HUG} web-page publicly available for everyone, and brought the HUG recommendations to the attention of the Herschel Science Team and the \ac{SGS}. 

Thus the \ac{HUG} was "by and for" the user community. The original membership was eight people drawn from the already existing \ac{KPOT} user community, later extended by four additional members drawn from the \ac{OT}1 community. \ac{HUG} members served for two years, and thus the membership was updated on a regular basis throughout, also including members drawn from the \ac{OT}2 community. \ac{NASA} had already somewhat earlier put in place their own \ac{NUP}, and the two committees not only collaborated by also had overlapping membership, the \ac{NUP} chair was a member of the \ac{HUG}. A summary of the \ac{HUG} membership can be found in Table~\ref{tab:Herschel_HUG}.

\begin{table}
\centering
\caption{A summary of the aggregate HUG membership, its composition varied over time (see text). The HUG Chair is included in the Full HUG, six out of the ten meetings held had a female chair. \%f is the female percentage.}
\begin{tabular}{cccccccccc}
\hline \noalign{\smallskip}
\vspace{3pt}
&\multicolumn{3}{c}{Full HUG} & \multicolumn{3}{c}{HUG Members} & \multicolumn{3}{c}{HUG Chairs}\\
&Male & Female & \%f & Male & Female & \%f & Male & Female & \%f  \\
\hline \noalign{\smallskip}
&10  &  7  &  41  &   9   &  6  &  40  &   1   &  1  &  50  \\
\hline \noalign{\smallskip}
\end{tabular}
\label{tab:Herschel_HUG}
\end{table}

\section{\textbf{Proposers}}
\label{sec:Herschel_Proposers}

The \ac{OT} phases of all three Herschel AOs were open to the worldwide community in a similar manner to \ac{AO}s from other ESA missions. There were no rules applicable to nationalities, and there were also no quotas or other kinds of limitations or guidelines in place with respect to the allocation of the available observing time. 
When submitting a proposal in response to an \ac{AO} a single \ac{PI} had to be named in addition to all the other co-investigators, all with affiliations. From the perspective of \ac{ESA} the proposal \ac{PI} was the contact point for the proposal for all kinds of communication, e.g., the allocation of time and information provision. The information provided by the \ac{PI}s allows an examination of the nationality distribution of the proposers' institutes. Table~\ref{tab:Herschel_ProposalNationalities} shows the number of proposals submitted and accepted covering all 1170 proposals. Accepted proposals are ones which were awarded any observing time at either priority, i.e. both p1 and p2 (cf. Sect.~\ref{sec:Herschel_ObservingTime} Observing Time).

It is immediately noticeable that just over 40\% of all submitted proposals were from \ac{PI}s located in institutes in the USA. This is slightly more than the sum of the five \ac{ESA} Member States with the largest numbers of proposals submitted; the same is true when it comes to numbers of accepted proposals. This illustrates one effect of having a fully open proposal procedures, another effect is the science return in terms of refereed publications ( Sect.~\ref{sec:ScienceProgrammePubs}).

\begin{table}
\centering
\caption{The countries where the Herschel PIs were located. The KPOT, OT1 and OT2 calls resulted in a total of 1170 submitted proposals with 644 (55.0\%) being accepted (allocated any observing time). The PIs were located at institutes in the following countries.}
\begin{tabular}{lcccc}
\hline \noalign{\smallskip}
\vspace{3pt}
Country	& No. Proposals	& Percentage  &    No.   &	Percentage \\
        &  Submitted    &  of Total   &	Accepted &	Accepted   \\
\hline \noalign{\smallskip}
BELGIUM	           &  13 &	1.1 &    7 &   53.8  \\
DENMARK            &   5 &  0.4 &    1 &   20.0  \\
FINLAND	           &   3 &	0.3 &    2 &   66.7  \\
FRANCE	           &  90 &	7.7 &   58 &   64.4  \\
GERMANY	           & 101 &  8.6 &	49 &   48.5  \\
GREECE	           &   3 &  0.3 &	 1 &   33.3  \\
HUNGARY            &  15 &  1.3 &    7 &   46.7  \\
IRELAND	           &   1 &  0.1 & 	 1 &  100  \\
ITALY              &  60 &	5.1 &	29 &   48.3  \\
NETHERLANDS	       &  69 &	5.9 &	41 &   59.4  \\
POLAND	           &   3 &	0.3 & 	 2 &   66.7  \\
PORTUGAL           &   2 &  0.2 &    2 &  100 \\
SPAIN              &  75 &  6.4 &   35 &   46.7  \\
SWEDEN             &  15 &  1.3 &	 8 &   53.3  \\
SWITZERLAND	       &  10 &	0.9 &	 5 &   50.0  \\
UNITED KINGDOM	   & 132 & 11.3 &	75 &   56.8  \\
ARMENIA            &   2 &  0.2 &    1 &   50.0  \\
AUSTRALIA          &  11 &	0.9 &	 7 &   63.6  \\
CANADA             &  31 &  2.6 &   18 &   58.1  \\
CHILE              &   1 &  0.1 &    1 &  100  \\
CHINA	           &   2 &  0.2 &	 2 &  100  \\
INDIA	           &   5 &	0.4 &	 2 &   40.0  \\
ISRAEL	           &   4 &	0.3 &	 4 &  100  \\
JAPAN	           &  29 &	2.5 &	12 &   41.4  \\
RUSSIAN FEDERATION &   2 &	0.2 &    0 &	0.0  \\
SLOVAKIA           &   1 &  0.1 &    0 &    0.0  \\
SOUTH AFRICA       &   2 &  0.2 &    1 &   50.0  \\
SOUTH KOREA        &   4 &  0.3 &    1 &   25.0  \\
TAIWAN      	   &   1 &	0.1 &	 1 &  100  \\
UNITED STATES      & 478 & 40.9 &  271 &   56.7  \\
\hline \noalign{\smallskip}
\end{tabular}
\label{tab:Herschel_ProposalNationalities}
\end{table}

\section{\textbf{Proposal Selection}}
\label{subsec:HerschelSelection}
\label{sec:Herschel_PropSel}

We examined the outcomes in terms of success based on gender for the \ac{OT} announcements (KPOT, OT1 and OT2) which, as discussed in Sect. (\ref{sec:Herschel_Proposers} were fully open to the worldwide community, without restrictions of any kind.  Of the 1170 proposals submitted in response to these \ac{AO}s, 644 (55.0\%) of the them were awarded {\it any} observing time, i.e. p1 or p2 time, and not necessarily all the time requested. 

In order to investigate success rates of proposals for male and female \ac{PI}s it was necessary to assign genders to the proposers since proposers were not asked to specify their gender (nor age, nor type of position held etc.). As with XMM-Newton and INTEGRAL, gender information for each proposer was assigned by the project scientist and \ac{SOC} staff using their knowledge of the community and through examining publicly-accessible web-based data in a similar way as for the \ac{HST} study by I.N.\ Reid ~\cite{2014Reid}. We appreciate that gender identity is more complex than a binary issue, however, no attempt can be made to assign genders other than male or female as this information is not readily available. It is possible that this process underestimates the number of female proposers, particularly those with non-Western names. However, this is likely be insignificant given the small fraction of non-Western proposers.

We first examined the number of proposals that were awarded any observing time. As listed in Table~\ref{tab:Herschel_AOgendernumbers} 828 proposals with male PIs and 342 proposals with female PIs were submitted from which 461 and 183 proposals were awarded any observing time. This gives success rates of 55.7\% and 53.5\% for proposals led by male and female PIs, respectively. This is a difference in favour of male PIs of 4.1\%. 

\begin{table}
\centering
\caption{Herschel OT proposals submitted with breakdown of proposals with male and female PIs and the number and percentages of proposals accepted as well as time and percentages awarded. KPOT proposals were not given priorities and all accepted proposals were counted as being priority p1.}
\setlength\tabcolsep{4.0pt}
\begin{tabular}{llccc}
\hline \noalign{\smallskip}
\vspace{3pt}
Priority & Parameter & \multicolumn{3}{c}{Proposal PI} \\
                   &    & All & Male & Female \\
\hline \noalign{\smallskip}
& Proposals Submitted & 1170 & 828 & 342 \\
\hline \noalign{\smallskip}           
p1, p2 & Number Accepted & 644 & 461 & 183 \\
 & Percentage Accepted & 55.0 & 55.7 & 53.5 \\
 & Ratio Male/Female & & \multicolumn{2}{c}{1.041} \\
p1 &      Number Accepted & 386 & 281 & 105 \\
& Percentage Accepted & 33.0 & 33.9 & 30.7 \\
& Ratio Male/Female & & \multicolumn{2}{c}{1.105} \\
\hline \noalign{\smallskip}
&Time Requested (hour) &  51,422 & 39,996 & 11,446 \\
\hline \noalign{\smallskip}
p1, p2 & Time Accepted (hour) & 19,751 & 15,336 & 4415 \\
& Time Accepted (\%) & 38.4 & 38.3 & 38.6 \\
& Ratio Male/Female & & \multicolumn{2}{c}{0.994} \\
p1 & Time Accepted (hour) & 13,934 & 11,464 & 2469 \\
& Time Accepted (\%) & 27.1 & 28.7 & 21.6 \\
& Ratio Male/Female & & \multicolumn{2}{c}{1.328} \\
\hline \noalign{\smallskip}
& Overall Male/Female Difference & & \multicolumn{2}{c}{\llap{$-$}1--33\%}\\
\hline \noalign{\smallskip}
\end{tabular}
\label{tab:Herschel_AOgendernumbers}
\end{table}

Proposal submissions and proposal acceptance are unlikely to be independent or random processes and for Poisson statistics to apply events need to be independent of each other. However, to illustrate the uncertainties that would apply if such statistics are applicable, we used square root uncertainties to update the success rates to be $(55.7 \pm 3.2)$\% and $(53.5 \pm 4.9$)\%  for male and female PI proposals, respectively. This is a difference in favour of males of ($4.1 \pm 11.3$)\%. We emphasise that these uncertainties are only given to illustrate the outcomes if Poisson statistics were to apply to the selection process. We note that actual uncertainties on the proposal numbers for each AO are zero. We further note that there are systematic uncertainties associated with this process due to misappropriated genders and incorrect \ac{PhD} "academic ages", but these are likely to be too small to significantly affect our results.

\begin{table}
\centering
\caption{The requested and allocated observing time for male and female Herschel proposers. The last two columns provide the percentage of the allocated as a fraction of the requested observing time for the genders.}
\setlength\tabcolsep{4.0pt}
\begin{tabular}{lcccccccccc}
\hline \noalign{\smallskip}
\vspace{3pt}
AO & \multicolumn{6}{c}{Time (hours)} & \multicolumn{4}{c}{Time (Percentage)}\\
      &  \multicolumn{2}{c}{Requested} & \multicolumn{2}{c}{Allocated (p1, p2)} & \multicolumn{2}{c}{Allocated (p1)}& \multicolumn{2}{c}{Allocated (p1, p2)} &
      \multicolumn{2}{c}{Allocated (p1)}\\
      &  Male & Female &  Male & Female & Male & Female & Male & Female & Male & Female\\
\hline \noalign{\smallskip}
KPOT & 15,729 &  2255 &	5001 &	 378 & 5001 & 378 &	31.8 &	16.8  & 31.8 & 16.8  \\
OT1  & 15,149 &	5823 &	4776 &	1801 & 3739 & 1238 & 31.5 &	30.9  & 24.7 & 21.3  \\
OT2  &  9118 &	3368 &	5559 &	2236 & 2725 & 853 &  61.0 &	66.4 & 29.9 & 25.3 \\
\textbf{Total} &\textbf{39,996} & \textbf{11,446} & \textbf{15,336} & \textbf{4415} & \textbf{11,464} & \textbf{2469} &  \textbf{38.4} & \textbf{38.7} & \textbf{28.7} & \textbf{21.6}\\
\hline \noalign{\smallskip}
\end{tabular}
\label{tab:Herschel_AOgendertime}
\end{table}

We examined the number of AO1 and AO2 proposals that were awarded p1 observing time (\ac{KPOT} proposals are counted as p1). As also listed in Table~\ref{tab:Herschel_AOgendernumbers} of the 828 proposals with male PIs and 342 proposals with female PIs that were submitted, 281 and 105 proposals were awarded p1 observing time. This gives success rates of 33.9\% and 30.7\% for proposals led by male and female PIs, respectively. This is a difference in favour of male PIs of 10.5\%.

We next examined the acceptance rates for the two genders using the awarded amounts of observing time. Many accepted Herschel proposals were allocated less observing time than requested.  If the allocation of observing time was handled differently for male and female \ac{PI}s, this will show as a difference in the relative amounts of time approved, compared to the number of proposals accepted. The results for each AO are summarised in Table~\ref{tab:Herschel_AOgendertime}. This shows the total times requested by proposals with male and female \ac{PI}s were 39,996~hour and 11,446~hour, respectively. A total of 15,336~hour and 4415~hour of p1+p2 observing time were awarded to proposals with male and female \ac{PI}s, respectively. This gives average success rates of 38.3\% for male \ac{PI}s and 38.6\% for female \ac{PI}s. This is almost equal with a small ($<$1\%) difference in favour of female \ac{PI}s. However, when only p1 time is considered a total of 11,464~hour and 2469~hour observing time were awarded to proposals with male and female \ac{PI}s, respectively. This gives average success rates of 28.7\% for male \ac{PI}s and 21.6\% for female \ac{PI}s, which amounts to a difference of 32.8\% in favour of male PIs. Initially we were intrigued by this situation, given all observing time there is negligible difference, but for p1 time only the situation is different. We investigated possible explanations for this. An inspection of Table~\ref{tab:Herschel_AOgendertime} indicates that the KPOT numbers stand out. In order to investigate further Table~\ref{tab:Herschel_AOgendernumbers2} was constructed, separating the KPOT results from the total (All) numbers, and comparing them directly with OT1+OT2 only (i.e. All-KPOT). 

Inspection of Table~\ref{tab:Herschel_AOgendernumbers2} reveals that for OT1+OT2 male PIs have an advantage in number of accepted proposals of ($5.4 \pm 1.6$)\%, but were awarded 3\% less p1+p2 observing time. As above, we used square root uncertainties to illustrate the uncertainties that would apply if such statistics are applicable. When OT1+OT2 proposals that were allocated p1 observing time are considered, male PIs have an advantage in numbers of ($15.5 \pm 15.2$)\%) and in allocated time of 17.6\%. However, for the KPOT on its own the differences are more significant, here male PIs have an possible advantage over female PIs of $(40.7 \pm 117.4$)\% in numbers, and 89.6\% in time. Initially this was a surprise, and the reason was not obvious. 

However, KPOT is "special" as it comprised of a small number of proposals, but with very substantial amounts of both requested and allocated observing time. We speculated that there could be "small number statistics" issues at play. To test this hypothesis we looked at what difference just one proposal being accepted or rejected could make. As shown in Table~\ref{tab:Herschel_AOgendernumbers2} the KPOT outcome was 19 accepted proposals with male PIs and 2 with female \ac{PI}s, out of 54 and 8 submitted proposals, respectively. Now, imagine the situation where the outcome would have been 18 and 3 instead. This would have resulted in ($33.3 \pm 9.1$)\% accepted proposals with male \ac{PI}s, and ($37.5 \pm 25.4$)\% with female \ac{PI}s, a female advantage of 11.1\% (rather than a male advantage of 40.7\%). Regarding allocated time, the average amount of time allocated to male \ac{PI}s was 263 hour per successful proposal and to female \ac{PI}s less at 189 hour. Imagine that the hypothetical third successful female \ac{PI} proposal would have been allocated 263 hour. This would have resulted in 4738 hour awarded to male \ac{PI}s, or 30.1\%, and 641 hour for female \ac{PI}s, or 28.4\%, a male advantage of 6.0\% (rather than 89.6\%), and in fact, smaller than the male advantage for OT1+OT2 p1 time (17.6\%). This discussion shows that "small number statistics" are likely to be important in understanding the origin of the KPOT "outlier"; certainly the results are consistent with this hypothesis. At the same time, we are aware that this does not prove that this hypothesis necessarily is the correct explanation. 

\begin{table}
\centering
\caption{Similar to Table~\ref{tab:Herschel_AOgendernumbers} but providing the same information for KPOT and OT1+OT2 proposals separately in order to display the large influence of the small in number but large in time KPOT proposals. See discussion in the text.}
\setlength\tabcolsep{4.0pt}
\begin{tabular}{llcccccc}
\hline \noalign{\smallskip}
\vspace{3pt}
Priority & Parameter & \multicolumn{6}{c}{Proposal PI} \\
                   &    & OT1+OT2 & Male & Female  & KPOT & Male & Female  \\
\hline \noalign{\smallskip}
& Proposals Submitted & 1108 & 774 & 334  & 62 & 54 & 8  \\
\hline \noalign{\smallskip}           
p1+p2 & Number Accepted & 623 & 442 & 181 & \dots & \dots & \dots  \\
 & Percentage Accepted & 56.2 & 57.1 & 54.2 & \dots & \dots & \dots \\
 & Ratio Male/Female & & \multicolumn{2}{c}{1.054}  & & \multicolumn{2}{c}{\dots} \\
p1    & Number Accepted & 386 & 281 & 105 & 21 & 19 & 2 \\
& Percentage Accepted & 34.8 & 36.3 & 31.4 & 33.9 & 35.2 & 25.0 \\
& Ratio Male/Female & & \multicolumn{2}{c}{1.155} & & \multicolumn{2}{c}{1.407} \\
\hline \noalign{\smallskip}
&Time Requested (hour) &  33,458 & 24,267 & 9191 & 17,984 & 15,729 & 2255 \\
\hline \noalign{\smallskip}
p1+p2 & Time Accepted (hour) & 14,372 & 10,335 & 4037 & \dots & \dots & \dots \\
& Time Accepted (\%) & 43.0 & 42.6 & 43.9 & \dots & \dots & \dots \\
& Ratio Male/Female & & \multicolumn{2}{c}{0.970} & & \multicolumn{2}{c}{\dots}\\
p1 & Time Accepted (hour) & 8555 & 6464 & 2091 & 5379 & 5001 & 378 \\
& Time Accepted (\%) & 25.7 & 26.8 & 22.8 & 29.9 & 31.8 & 16.8 \\
& Ratio Male/Female & & \multicolumn{2}{c}{1.176} & & \multicolumn{2}{c}{1.896} \\
\hline \noalign{\smallskip}
& Overall Male/Female Difference & & \multicolumn{2}{c}{\llap{$-$}3--18\%} & & \multicolumn{2}{c}{41--90\%} \\
\hline \noalign{\smallskip}
\end{tabular}
\label{tab:Herschel_AOgendernumbers2}
\end{table}

The above discussion still leaves the male advantages for the OT1 and OT2 AOs. Since the female proposer population is expected to be less senior than the male one and thus on average less experienced, we have investigated the experience of the populations of male and female proposer PIs. In order to determine the "academic age" of proposers we used the difference between the year that their \ac{PhD} was awarded and the year that an AO to which they submitted a proposal was issued. Thus a proposer who submits multiple proposals to the same AO will be counted multiple times in the same "academic year", whilst one who submits multiple proposals to different AOs will be counted in different "academic years". 

The year a proposer obtained their \ac{PhD} (or equivalent) was determined for 96.4\% of the proposals by searching the internet, particularly sites such as the \ac{ADS}, LinkedIn, the Astronomy Genealogy Project (\url{astrogen.aas.org}), \url{ORCID.org}, IEEE Xplore (\url{https://ieeexplore.ieee.org/Xplore/home.jsp}), and, for French theses \url{https://www.theses.fr}. Some proposers were contacted and provided their \ac{PhD} dates by email. The proposers for which the \ac{PhD} could not be found are often retired, deceased, or have left astronomy for some other reason. For \ac{PhD} students who had not yet completed their degrees, the expected year of submission was used. For the small number of proposers who did not have a \ac{PhD} and were not enrolled in a \ac{PhD} programme, their dates were assumed to be arbitrarily far in the future. For some senior scientists in Italian institutes who do not have a PhD, a date three years after they obtained their Laurea was used. It should be noted that using the year of \ac{PhD} to indicate the seniority does not take into account possible time spent outside of astronomy such as a career elsewhere. 

\begin{figure*}[ht]
\centering
\includegraphics[width=0.96\textwidth,angle=0]{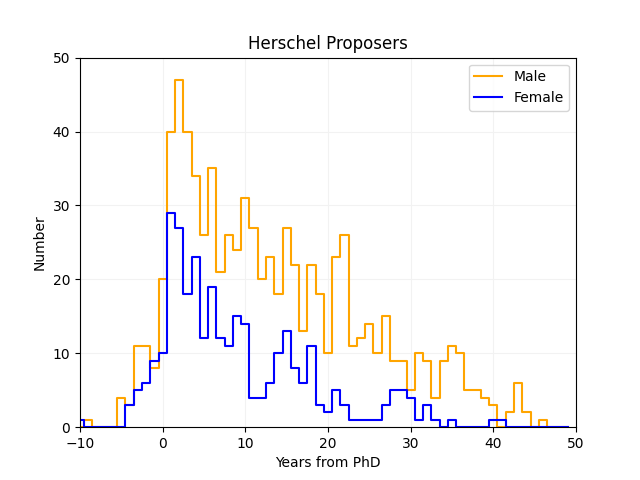}
\caption{The "academic age" (years since PhD) for male and female Herschel PIs. Male PIs have a mean "academic age" of 13.5 years compared to 8.9 years for female PIs.}
\label{fig:Herschel_ProposalAge}    
\end{figure*}

Figure~\ref{fig:Herschel_ProposalAge} shows the numbers of male and female Herschel \ac{PI}s who obtained their \ac{PhD}s or equivalent in one year bins between 10 years before the \ac{PhD} date to 50 years after.  Male PIs have a mean "academic age" of 13.5 years post-PhD compared to that of the female PIs of 8.9 years. The mean "academic age" of all \ac{PI}s is 12.2 years. A two-sample Anderson-Darling Test (e.g., \cite{2006Babu}) shows that the hypothesis that both samples come from the same underlying population can be rejected at $>$99\% confidence. 

\begin{figure*}[ht]
\centering
\includegraphics[width=0.96\textwidth,angle=0]{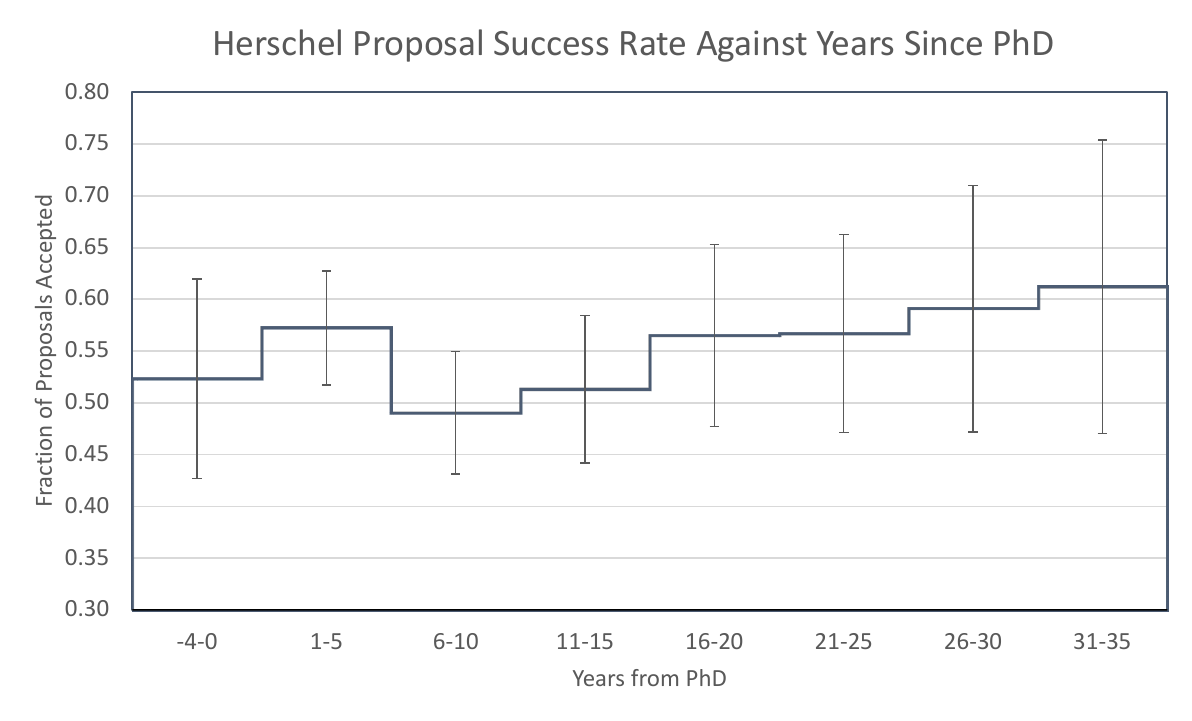}
\caption{The acceptance fraction of all Herschel PIs with year of their PhD. The indicative error bars indicate 1$\sigma$ standard deviations assuming that the number of proposers in each age bin follows a Poisson distribution.}
\label{fig:Herschel_ProposalPhD}    
\end{figure*}

We investigated the acceptance rates of proposals against \ac{PhD} year. This is shown plotted for years $-$4 to 35~years after \ac{PhD} in five year bins in Fig.~\ref{fig:Herschel_ProposalPhD}. To the eye it may look like a slow increase in acceptance rates for above 10 years of academic age, however, the error bars are large and a least squares fit to the un-binned data gives an intercept of ($52.5 \pm 4.3$) \% and a gradient of ($0.03 \pm 0.29$)\% year$^{-1}$ assuming that the uncertainties are the square root of the number of people in each bin. The value of R$^2$ is 0.42. The uncertainties are large and the gradient is consistent with no change in proposal acceptance rate with "academic age". This is in contrast to the results from INTEGRAL and XMM-Newton which both show stronger evidence for increasing proposal success rates with "academic age". 

\section{\textbf{Comparison with the Infrared Space Observatory (ISO)}}
\label{sec:Herschel_ISO}

The study of Herschel AOs and their outcomes is basically a snapshot from around the year 2010. This is in contrast to the $\sim$20 year spans of AOs for XMM-Newton and INTEGRAL. However, as alluded to in Sect.~\ref{sec:Herschel_ScienceProd}, the Herschel community has a connection with earlier missions, notably the ESA \ac{ISO} mission~\cite{1996Kessler}. 

\ac{ISO} predates the Horizon 2000 programme~\cite{1984ESA}. It was selected in 1983 based on a proposal from 1979, and launched on 17 November 1995. \ac{ISO} was the world's first true orbiting infrared space observatory, providing astronomers with unprecedented sensitivity and observation capabilities for a detailed exploration of the universe at infrared wavelengths. Equipped with four state-of-the-art instruments, the \ac{ISOCAM}, the \ac{ISOPHOT}, the \ac{SWS} and the \ac{LWS} housed in a superfluid helium cryostat (the forerunner of the Herschel cryogenic system) it probed the sky in a range of photometric, polarimetric and spectroscopic modes in the spectral range 2.5--240\,$\mu$m. The routine operational phase lasted until April 1998 during which time \ac{ISO} made some 30,000 observations.

For \ac{ISO} we do not have the same amount of data pertaining in particular to the statistics for proposals and observing time; at the time this was handled in a much more "manual" fashion compared to later missions. There were two \ac{AO}s for observing proposals, featuring \ac{GT} and \ac{OT} as for other observatories, a pre-launch call in 1994, and a supplemental call in 1996. We could only find information on the total number of accepted proposals in the two calls which can be found in Table~\ref{tab:Herschel_ISO}. Unfortunately, information on the number of submitted proposals could not be found.

\begin{table}[ht]
\centering
\caption{The number of accepted proposals for the ISO GT and OT calls and the corresponding numbers with female PIs. Information on the number of submitted proposals is unfortunately missing. \%(f)PI is the percentage of accepted proposals with a female PI.}
\setlength\tabcolsep{5.0pt}
\begin{tabular}{cccccccccc}
\hline \noalign{\smallskip}
\vspace{3pt}
&\multicolumn{3}{c}{ISO GT} & \multicolumn{3}{c}{ISO OT} & \multicolumn{3}{c}{ISO total}\\
& No.props & No.(f)PI & \%(f)PI & No.props & No. (f)PI & \%(f)PI & No. props & No. (f)PI & \%(f)PI  \\
\hline \noalign{\smallskip}
& 168  & 21  & 12.5  & 911 & 141 & 15.5 & 1079 & 162 & 15.0 \\
\hline \noalign{\smallskip}
\end{tabular}
\label{tab:Herschel_ISO}
\end{table}

The fraction of all accepted \ac{ISO} proposals with a female PI is 15.0\%. For the OT only it is 15.5\% which can be compared with the same figure for Herschel (from Table~\ref{tab:Herschel_AOgendernumbers}) with a total of 644 OT proposals accepted out of which 183 had a female PI, or 28.4\%. For the Herschel p1 proposals (only) the numbers are a total of 386 OT proposals accepted out of which 105 had a female PI, or 27.2\%. This means that the fraction of successful female PIs increased by a factor of 1.83 (all) or 1.75 (p1 only) between around 1995 for ISO and about 15 years later for Herschel. This may reflect an increase in the fraction of female (infrared) astronomers in the community by about 80\% in the time of Herschel compared to the time of ISO.

To test this hypothesis the fraction of female astronomers in the community over time are needed. We have had limited success finding this data and we extracted results between mid 2011 to late 2023 from the \ac{IAU}'s membership statistics (Digital Public Data) (see Sect.~\ref{sec:ScienceProgramme}).  In mid-2011 the fraction of female \ac{IAU} members was 17.5\%. This is below the fraction of successful Herschel proposals with female PIs of $\sim$28\% in the preceding years which is a factor $\sim$1.6 higher. However, without the fraction of female astronomers in the community in around 1995 it is not possible to compare the corresponding \ac{ISO} number of $\sim$15\% with the fraction of female IAU members at the time. If the corresponding \ac{ISO} number also would be a factor of $\sim$1.6 higher it would have been tempting to ascribe the increase in the fraction of female PIs from ISO to Herschel simply resulting from the increase in the fraction of female astronomers overall. Aside from the fact that we lack the required numbers, this could of course be an overly simplistic explanation. 

Independently, we have received information concerning the membership of the \ac{AAS}, also incomplete, but indicating that the fraction of female members may have increased by a factor approximately in the range 1.45--1.65 between 1995 and 2010. If the actual number is anywhere in this range it is thus somewhat smaller than the $\sim$1.8 ratio of the fractions of successful proposals with female \ac{PI}s in Herschel versus \ac{ISO}. This would thus go part of the way but needing additional components to fully explain the evolution observed, such as e.g., an increase in the fraction of infrared/sub-millimetre astronomers in the total population of astronomers, which would not at all be unthinkable in this time-frame with all the activities in the field, in space and on the ground.

There continue to be $\sim$30 refereed ISO publications per year and by the end of 2021 the total number is approaching 2100, another example of longevity. An interesting aspect of the \ac{ISO} publication curve is what occurred in the "Herschel time", when the annual number of \ac{ISO} publications approximately doubled in the early years of Herschel 2009--2011, and was elevated for another three years. This is hypothesised as a "Herschel effect", more people may have been using \ac{ISO} data with respect to planning and proposing for Herschel, and in interpreting Herschel data.  Furthermore, although not easily visible in the number of Herschel publications curve as an increase similar to that of \ac{ISO}, several of the early \ac{ALMA} observations had Herschel connections, using Herschel data, and generating Herschel publications. It appears that "next generation" missions may enhance the value of the data from earlier missions, generating additional publications in the process.

\section{\textbf{Discussion}}
\label{sec:Herschel_Discussion}

Herschel is a highly successful mission that scores well in various "success metrics" as already pointed out in Sect.~\ref{sec:Herschel_ScienceProd}. As described and shown in Fig.~\ref{fig:Herschel_ESAledobs10yrpubs} Herschel's productivity in terms of publications began early in the mission. This is not something that "happened" but was due to careful preparations in order to be able to use the observatory efficiently, since its short lifetime was consumable (cryogen) limited, thus the onus was on efficient operations from the very beginning.

The introduction of the \ac{SDP} was generally considered a big success; it validated the actual science performance in various respects, it provided real-life performance numbers in time already for the first in-flight \ac{AO} and it provided material used for outreach.

As already pointed out Herschel provides only snapshots when it comes to studies of the proposal and time allocation processes and their evolution. There are two aspects that we find interesting. The first is that, if we neglect the KPOT selection, there is a probable male advantage of ($-$3--18\%) in these processes (see Table~\ref{tab:Herschel_AOgendernumbers2}, which is similar to that found for XMM-Newton and INTEGRAL (see \cite{2024XMM} and \cite{2024INTEGRAL}). The second is, indeed, the very different situation for the KPOT selection which at first sight appears to have been strongly in favour of male PIs (Table~\ref{tab:Herschel_AOgendernumbers2}). However, although it cannot be positively confirmed, it is possible (see the discussion in Sect.~\ref{sec:Herschel_PropSel}) that the outcome has been strongly influenced by "small number statistics".

Supplemented by what we have in terms of proposal statistics from \ac{ISO} in around 1995 (about 15 years before Herschel), we find that the fraction of successful proposals with female PIs is about 80\% higher for Herschel than for ISO. This appears to be a larger increase than the increase in the female fraction of the overall astronomical community in the same time, but more statistical data would be needed to affirm this tentative conclusion.

\noindent
\clearpage
\textbf{Acknowledgements.}
\small{We thank the Herschel \ac{HSC} staff for their assistance.}

\clearpage
\printbibliography[keyword=Herschel,heading=subbibliography,title={\textbf{References}}]

@BOOK{1984ESA,
       author = {},
      	title = "{European Space Science -- Horizon 2000}",   
         year = {1984},
      address = {SP--1070},
      editor  = {},
    publisher = {European Space Agency}        
}

@ARTICLE{2001jansen,
   author = {{Jansen}, F. and {Lumb}, D. and {Altieri}, B. and {Clavel}, J. and 
	{Ehle}, M. and {Erd}, C. and {Gabriel}, C. and {Guainazzi}, M. and 
	{Gondoin}, P. and {Much}, R. and {Munoz}, R. and {Santos}, M. and 
	{Schartel}, N. and {Texier}, D. and {Vacanti}, G.},
    title = "{XMM-Newton observatory. I. The spacecraft and operations}",
  journal = {\aap},
     year = 2001,
    month = jan,
   volume = 365,
    pages = {L1-L6}
}

@ARTICLE{2014Reid,
       author = {{Reid}, I. N.},
        title = "{Gender-Correlated Systematics in HST Proposal Selection}",
      journal = {\pasp},
         year = 2014,
        month = oct,
       volume = {126},
       number = {944},
        pages = {923}
}

@ARTICLE{2006Babu,
        author = {{Babu}, G. J. and {Feigelson}, E. D.},
        title = "{Goodness-of-fit and all that!}",
        journal = {ASP Conf.},
        year = 2006,
        volume = {351},
        pages = {127}
}

@BOOK{2017HerschelY,
       author = {{Minier,} V. and {Bonnet,} R.-M. and {Bontems,} V. and {T. de Graauw,} and {M. Griffin,} and {F. Helmich,} and {G. Pilbratt,} and {S. Volonte}},
      	title = "{Inventing a Space Mission - The Story of the Herschel Space Observatory}",   
         year = {2017},
      address = {ISSI Scientific Report Series, volume 14},
      editor  = {},
    publisher = {Springer}        
}

@BOOK{2019HerschelSAL,
       author = {{Clark,} S. and {Pilbratt, G.}},
      	title = "{Herschel - Science and Legacy}", year = {2019},
      address = {ESA Brochure},
      editor  = {{O'Flaherty} K. O. and {Pilbratt} G.},
    publisher = {ESA}        
}

@BOOK{1997Grenoble,
       author = {},
      	title = "{The Far Infrared and Submillimetre Universe}", 
      	year = {1997},
      address = {ESA SP-401},
      editor  = {{A. Wilson}},
    publisher = {ESA}        
}

@BOOK{2001Toledo,
       author = {},
      	title = "{The Promise of the Herschel Space Observatory}", 
      	year = {2001},
      address = {ESA SP-460},
      editor  = {{G. L. Pilbratt,} {J. Cernicharo,} {A. M. Heras,} {T. Prusti,} and {R. Harris}},
    publisher = {ESA}        
}

@ARTICLE{2010Pilbratt,
       author = {{Pilbratt}, G.~L. and {Riedinger}, J.~R. and {Passvogel}, T. and {Crone}, G. and {Doyle}, D. and {Gageur}, U. and {Heras}, A.~M. and {Jewell}, C. and {Metcalfe}, L. and {Ott}, S. and {Schmidt}, M.},
        title = "{Herschel Space Observatory. An ESA facility for far-infrared and submillimetre astronomy}",
      journal = {\aap},
         year = 2010,
        month = jul,
       volume = {518},
        pages = {L1}
}

@ARTICLE{1996Kessler,
       author = {{Kessler}, M.~F. and {Steinz}, J.~A. and {Anderegg}, M.~E. and {Clavel}, J. and {Drechsel}, G. and {Estaria}, P. and {Faelker}, J. and {Riedinger}, J.~R. and {Robson}, A. and {Taylor}, B.~G. and {Xim{\'e}nez de Ferr{\'a}n}, S.},
        title = "{The Infrared Space Observatory (ISO) mission.}",
      journal = {\aap},
         year = 1996,
        month = nov,
       volume = {315},
       number = {2},
        pages = {L27}
}

@ARTICLE{2009Doyle,
       author = {{Doyle}, D. and {Pilbratt}, G. and {Tauber}, J.},
        title = "{The Herschel and Planck Space Telescopes}",
      journal = {\ieee},
         year = 2009,
        month = aug,
       volume = {97},
        pages = {1043}
}

@ARTICLE{2010Poglitsch,
       author = {{Poglitsch}, A. and {Waelkens}, C. and {Geis}, N. and {Feuchtgruber}, H. and {Vandenbussche}, B. and {Rodriguez}, L. and {Krause}, O. and {Renotte}, E. and {van Hoof}, C. and {Saraceno}, P. and {Cepa}, J. and {Kerschbaum}, F. and {Agn{\`e}se}, P. and {Ali}, B. and {Altieri}, B. and {Andreani}, P. and {Augueres}, J. -L. and {Balog}, Z. and {Barl}, L. and {Bauer}, O.~H. and {Belbachir}, N. and {Benedettini}, M. and {Billot}, N. and {Boulade}, O. and {Bischof}, H. and {Blommaert}, J. and {Callut}, E. and {Cara}, C. and {Cerulli}, R. and {Cesarsky}, D. and {Contursi}, A. and {Creten}, Y. and {De Meester}, W. and {Doublier}, V. and {Doumayrou}, E. and {Duband}, L. and {Exter}, K. and {Genzel}, R. and {Gillis}, J. -M. and {Gr{\"o}zinger}, U. and {Henning}, T. and {Herreros}, J. and {Huygen}, R. and {Inguscio}, M. and {Jakob}, G. and {Jamar}, C. and {Jean}, C. and {de Jong}, J. and {Katterloher}, R. and {Kiss}, C. and {Klaas}, U. and {Lemke}, D. and {Lutz}, D. and {Madden}, S. and {Marquet}, B. and {Martignac}, J. and {Mazy}, A. and {Merken}, P. and {Montfort}, F. and {Morbidelli}, L. and {M{\"u}ller}, T. and {Nielbock}, M. and {Okumura}, K. and {Orfei}, R. and {Ottensamer}, R. and {Pezzuto}, S. and {Popesso}, P. and {Putzeys}, J. and {Regibo}, S. and {Reveret}, V. and {Royer}, P. and {Sauvage}, M. and {Schreiber}, J. and {Stegmaier}, J. and {Schmitt}, D. and {Schubert}, J. and {Sturm}, E. and {Thiel}, M. and {Tofani}, G. and {Vavrek}, R. and {Wetzstein}, M. and {Wieprecht}, E. and {Wiezorrek}, E.},
        title = "{The Photodetector Array Camera and Spectrometer (PACS) on the Herschel Space Observatory}",
      journal = {\aap},
         year = 2010,
        month = jul,
       volume = {518},
        pages = {L2}
}

@ARTICLE{2010Griffin,
       author = {{Griffin}, M.~J. and {Abergel}, A. and {Abreu}, A. and {Ade}, P.~A.~R. and {Andr{\'e}}, P. and {Augueres}, J. -L. and {Babbedge}, T. and {Bae}, Y. and {Baillie}, T. and {Baluteau}, J. -P. and {Barlow}, M.~J. and {Bendo}, G. and {Benielli}, D. and {Bock}, J.~J. and {Bonhomme}, P. and {Brisbin}, D. and {Brockley-Blatt}, C. and {Caldwell}, M. and {Cara}, C. and {Castro-Rodriguez}, N. and {Cerulli}, R. and {Chanial}, P. and {Chen}, S. and {Clark}, E. and {Clements}, D.~L. and {Clerc}, L. and {Coker}, J. and {Communal}, D. and {Conversi}, L. and {Cox}, P. and {Crumb}, D. and {Cunningham}, C. and {Daly}, F. and {Davis}, G.~R. and {de Antoni}, P. and {Delderfield}, J. and {Devin}, N. and {di Giorgio}, A. and {Didschuns}, I. and {Dohlen}, K. and {Donati}, M. and {Dowell}, A. and {Dowell}, C.~D. and {Duband}, L. and {Dumaye}, L. and {Emery}, R.~J. and {Ferlet}, M. and {Ferrand}, D. and {Fontignie}, J. and {Fox}, M. and {Franceschini}, A. and {Frerking}, M. and {Fulton}, T. and {Garcia}, J. and {Gastaud}, R. and {Gear}, W.~K. and {Glenn}, J. and {Goizel}, A. and {Griffin}, D.~K. and {Grundy}, T. and {Guest}, S. and {Guillemet}, L. and {Hargrave}, P.~C. and {Harwit}, M. and {Hastings}, P. and {Hatziminaoglou}, E. and {Herman}, M. and {Hinde}, B. and {Hristov}, V. and {Huang}, M. and {Imhof}, P. and {Isaak}, K.~J. and {Israelsson}, U. and {Ivison}, R.~J. and {Jennings}, D. and {Kiernan}, B. and {King}, K.~J. and {Lange}, A.~E. and {Latter}, W. and {Laurent}, G. and {Laurent}, P. and {Leeks}, S.~J. and {Lellouch}, E. and {Levenson}, L. and {Li}, B. and {Li}, J. and {Lilienthal}, J. and {Lim}, T. and {Liu}, S.~J. and {Lu}, N. and {Madden}, S. and {Mainetti}, G. and {Marliani}, P. and {McKay}, D. and {Mercier}, K. and {Molinari}, S. and {Morris}, H. and {Moseley}, H. and {Mulder}, J. and {Mur}, M. and {Naylor}, D.~A. and {Nguyen}, H. and {O'Halloran}, B. and {Oliver}, S. and {Olofsson}, G. and {Olofsson}, H. -G. and {Orfei}, R. and {Page}, M.~J. and {Pain}, I. and {Panuzzo}, P. and {Papageorgiou}, A. and {Parks}, G. and {Parr-Burman}, P. and {Pearce}, A. and {Pearson}, C. and {P{\'e}rez-Fournon}, I. and {Pinsard}, F. and {Pisano}, G. and {Podosek}, J. and {Pohlen}, M. and {Polehampton}, E.~T. and {Pouliquen}, D. and {Rigopoulou}, D. and {Rizzo}, D. and {Roseboom}, I.~G. and {Roussel}, H. and {Rowan-Robinson}, M. and {Rownd}, B. and {Saraceno}, P. and {Sauvage}, M. and {Savage}, R. and {Savini}, G. and {Sawyer}, E. and {Scharmberg}, C. and {Schmitt}, D. and {Schneider}, N. and {Schulz}, B. and {Schwartz}, A. and {Shafer}, R. and {Shupe}, D.~L. and {Sibthorpe}, B. and {Sidher}, S. and {Smith}, A. and {Smith}, A.~J. and {Smith}, D. and {Spencer}, L. and {Stobie}, B. and {Sudiwala}, R. and {Sukhatme}, K. and {Surace}, C. and {Stevens}, J.~A. and {Swinyard}, B.~M. and {Trichas}, M. and {Tourette}, T. and {Triou}, H. and {Tseng}, S. and {Tucker}, C. and {Turner}, A. and {Vaccari}, M. and {Valtchanov}, I. and {Vigroux}, L. and {Virique}, E. and {Voellmer}, G. and {Walker}, H. and {Ward}, R. and {Waskett}, T. and {Weilert}, M. and {Wesson}, R. and {White}, G.~J. and {Whitehouse}, N. and {Wilson}, C.~D. and {Winter}, B. and {Woodcraft}, A.~L. and {Wright}, G.~S. and {Xu}, C.~K. and {Zavagno}, A. and {Zemcov}, M. and {Zhang}, L. and {Zonca}, E.},
        title = "{The Herschel-SPIRE instrument and its in-flight performance}",
      journal = {\aap},
         year = 2010,
        month = jul,
       volume = {518},
        pages = {L3}
}

@ARTICLE{2010deGraauw,
       author = {{de Graauw}, Th. and {Helmich}, F.~P. and {Phillips}, T.~G. and {Stutzki}, J. and {Caux}, E. and {Whyborn}, N.~D. and {Dieleman}, P. and {Roelfsema}, P.~R. and {Aarts}, H. and {Assendorp}, R. and {Bachiller}, R. and {Baechtold}, W. and {Barcia}, A. and {Beintema}, D.~A. and {Belitsky}, V. and {Benz}, A.~O. and {Bieber}, R. and {Boogert}, A. and {Borys}, C. and {Bumble}, B. and {Ca{\"\i}s}, P. and {Caris}, M. and {Cerulli-Irelli}, P. and {Chattopadhyay}, G. and {Cherednichenko}, S. and {Ciechanowicz}, M. and {Coeur-Joly}, O. and {Comito}, C. and {Cros}, A. and {de Jonge}, A. and {de Lange}, G. and {Delforges}, B. and {Delorme}, Y. and {den Boggende}, T. and {Desbat}, J. -M. and {Diez-Gonz{\'a}lez}, C. and {di Giorgio}, A.~M. and {Dubbeldam}, L. and {Edwards}, K. and {Eggens}, M. and {Erickson}, N. and {Evers}, J. and {Fich}, M. and {Finn}, T. and {Franke}, B. and {Gaier}, T. and {Gal}, C. and {Gao}, J.~R. and {Gallego}, J. -D. and {Gauffre}, S. and {Gill}, J.~J. and {Glenz}, S. and {Golstein}, H. and {Goulooze}, H. and {Gunsing}, T. and {G{\"u}sten}, R. and {Hartogh}, P. and {Hatch}, W.~A. and {Higgins}, R. and {Honingh}, E.~C. and {Huisman}, R. and {Jackson}, B.~D. and {Jacobs}, H. and {Jacobs}, K. and {Jarchow}, C. and {Javadi}, H. and {Jellema}, W. and {Justen}, M. and {Karpov}, A. and {Kasemann}, C. and {Kawamura}, J. and {Keizer}, G. and {Kester}, D. and {Klapwijk}, T.~M. and {Klein}, Th. and {Kollberg}, E. and {Kooi}, J. and {Kooiman}, P. -P. and {Kopf}, B. and {Krause}, M. and {Krieg}, J. -M. and {Kramer}, C. and {Kruizenga}, B. and {Kuhn}, T. and {Laauwen}, W. and {Lai}, R. and {Larsson}, B. and {Leduc}, H.~G. and {Leinz}, C. and {Lin}, R.~H. and {Liseau}, R. and {Liu}, G.~S. and {Loose}, A. and {L{\'o}pez-Fernandez}, I. and {Lord}, S. and {Luinge}, W. and {Marston}, A. and {Mart{\'\i}n-Pintado}, J. and {Maestrini}, A. and {Maiwald}, F.~W. and {McCoey}, C. and {Mehdi}, I. and {Megej}, A. and {Melchior}, M. and {Meinsma}, L. and {Merkel}, H. and {Michalska}, M. and {Monstein}, C. and {Moratschke}, D. and {Morris}, P. and {Muller}, H. and {Murphy}, J.~A. and {Naber}, A. and {Natale}, E. and {Nowosielski}, W. and {Nuzzolo}, F. and {Olberg}, M. and {Olbrich}, M. and {Orfei}, R. and {Orleanski}, P. and {Ossenkopf}, V. and {Peacock}, T. and {Pearson}, J.~C. and {Peron}, I. and {Phillip-May}, S. and {Piazzo}, L. and {Planesas}, P. and {Rataj}, M. and {Ravera}, L. and {Risacher}, C. and {Salez}, M. and {Samoska}, L.~A. and {Saraceno}, P. and {Schieder}, R. and {Schlecht}, E. and {Schl{\"o}der}, F. and {Schm{\"u}lling}, F. and {Schultz}, M. and {Schuster}, K. and {Siebertz}, O. and {Smit}, H. and {Szczerba}, R. and {Shipman}, R. and {Steinmetz}, E. and {Stern}, J.~A. and {Stokroos}, M. and {Teipen}, R. and {Teyssier}, D. and {Tils}, T. and {Trappe}, N. and {van Baaren}, C. and {van Leeuwen}, B. -J. and {van de Stadt}, H. and {Visser}, H. and {Wildeman}, K.~J. and {Wafelbakker}, C.~K. and {Ward}, J.~S. and {Wesselius}, P. and {Wild}, W. and {Wulff}, S. and {Wunsch}, H. -J. and {Tielens}, X. and {Zaal}, P. and {Zirath}, H. and {Zmuidzinas}, J. and {Zwart}, F.},
        title = "{The Herschel-Heterodyne Instrument for the Far-Infrared (HIFI)}",
      journal = {\aap},
         year = 2010,
        month = jul,
       volume = {518},
        pages = {L6}
}

@ARTICLE{2009Riedinger,
       author = {{Riedinger}, J.},
        title = "{A First in Astrophysics Missions. The making of the Herschel Science Ground Segment}",
      journal = {ESA Bulletin},
         year = 2009,
        month = aug,
       volume = {139},
        pages = {12}
}

@INPROCEEDINGS{2010Ott,
       author = {{Ott}, S.},
        title = "{The Herschel Data Processing System - HIPE and Pipelines - Up and Running Since the Start of the Mission}",
    booktitle = {Astronomical Data Analysis Software and Systems XIX},
         year = 2010,
       editor = {{Mizumoto}, Y. and {Morita}, K.-I. and {Ohishi}, M.},
       series = {ASP Conference Series, Vol. 434},
        pages = {139}       
}

@INPROCEEDINGS{2013Ott,
       author = {{Ott}, S.},
        title = "{Herschel Data Processing Development: 10 Years After}",
    booktitle = {Astronomical Data Analysis Software and Systems XXII},
         year = 2013,
       editor = {{Friedel}, D.~N.},
       series = {ASP Conference Series, Vol. 475},
        pages = {197}       
}

@ARTICLE{2016Valiante,
       author = {{Valiante}, E. and {Smith}, M.~W.~L. and {Eales}, S. and {et al.}},
        title = "{The Herschel-ATLAS data release 1 – I. Maps, catalogues and number counts}",
      journal = {\mnras},
         year = 2016,
        month = jul,
       volume = {462},
        pages = {3146}
}

@ARTICLE{2018Maddox,
       author = {{Maddox}, S.~J. and {Valiante}, E. and {Cigan}, P. and {Dunne}, L. and {Eales}, S. and {et al.}},
        title = "{The Herschel-ATLAS Data Release 2. Paper II. Catalogs of Far-infrared and Submillimeter Sources in the Fields at the South and North Galactic Poles}",
      journal = {\apjs},
         year = 2018,
        month = jun,
       volume = {236},
        pages = {30}
}

@ARTICLE{2011Rodighiero,
        author = {{Rodighiero}, G. and {Daddi}, E. and {Baronchelli}, I. and {et al.}},
        title = "{The Lesser Role of Starbursts in Star Formation at z=2}",
      journal = {\apjl},
         year = 2011,
        month = oct,
       volume = {739},
        pages = {L40}
}

@ARTICLE{2018Elbaz,
        author = {{Elbaz}, D. and {Leiton}, R. and {Nagar}, N. and {et al.}},
        title = "{Starbursts in and out of the star-formation main sequence}",
      journal = {\aap},
         year = 2018,
       volume = {616},
        pages = {A110}
}

@ARTICLE{2011Elbaz,
        author = {{Elbaz}, D. and {Dickinson}, M. and {Hwang}, H.~S. and {et al.}},
        title = "{GOODS–Herschel: an infrared main sequence for star-forming galaxies}",
      journal = {\aap},
         year = 2011,
       volume = {533},
        pages = {A119}
}

@ARTICLE{2010Dye,
        author = {{Dye}, S. and {Dunne}, L. and {Eales}, S. and {et al.}},
        title = "{Herschel-ATLAS: Evolution of the 250 $\mu$m luminosity function out to z = 0.5}",
      journal = {\aap},
         year = 2010,
       volume = {518},
        pages = {L10}
}

@ARTICLE{2013Gruppioni,
        author = {{Gruppioni}, C. and {Pozzi}, F. and {Rodighiero}, G. and {et al.}},
        title = "{The Herschel PEP/HerMES luminosity function – I. Probing the evolution of PACS selected Galaxies to z $\simeq$ 4}",
      journal = {\mnras},
         year = 2013,
       volume = {432},
       month = apr,
        pages = {23}
}

@ARTICLE{2011Dunne,
        author = {{Dunne}, L. and {Gomez}, H.~L. and {da Cunha}, E. and {et al.}},
        title = "{Herschel-ATLAS: rapid evolution of dust in galaxies over the last 5 billion years}",
      journal = {\mnras},
         year = 2011,
       volume = {417},
        pages = {1510}
}

@ARTICLE{2014Lutz,
        author = {{Lutz}, D.},
        title = "{Far-Infrared Surveys of Galaxy Evolution}",
      journal = {Ann. Rev. Astron. Astrophys.},
         year = 2014,
       volume = {52},
        pages = {373}
}

@ARTICLE{2013Riechers,
        author = {{Riechers}, D.~A. and {Bradford},  C.~M. and {Clements}, D.~L. and {et al.}},
        title = "{A dust-obscured massive maximum-starburst galaxy at a redshift of 6.34}",
      journal = {\nat},
         year = 2013,
         month = apr,
       volume = {496},
        pages = {329}
}

@ARTICLE{2014Dowell,
        author = {{Dowell}, C.~D. and {Conley}, A. and {Glenn}, J. and {et al.}},
        title = "{HerMES: Candidate High-redshift Galaxies Discovered with Herschel/SPIRE}",
      journal = {\apj},
         year = 2014,
         month = jan,
       volume = {780},
        pages = {75}
}

@ARTICLE{2016Ivison,
        author = {{Ivison}, R.~J. and {Lewis},  A.~J.~R. and {Weiss}, A. and {et al.}},
        title = "{The Space Density of Luminous Dusty Star-forming Galaxies at z>4: SCUBA-2 and LABOCA Imaging of Ultrared Galaxies from Herschel-ATLAS}",
      journal = {\apj},
         year = 2016,
         month = nov,
       volume = {832},
        pages = {78}
}

@ARTICLE{2017Riechers,
        author = {{Riechers}, D.~A. and {Leung}, T.~K.~D. and {Ivison}, R.~J. and {et al.}},
        title = "{Rise of the Titans: A Dusty, Hyper-luminous “870 $\mu$m Riser” Galaxy at z $\sim$ 6}",
      journal = {\apj},
         year = 2017,
         month = nov,
       volume = {850},
        pages = {1}
}

@ARTICLE{2010Fischer,
        author = {{Fischer}, J. and {Sturm}, E. and {Gonz\'{a}lez-Alfonso}, E. and {et al.}},
        title = "{Herschel-PACS spectroscopic diagnostics of local ULIRGs: Conditions and kinematics in Markarian 231}",
      journal = {\aap},
         year = 2010,
         month = jul,
       volume = {518},
        pages = {L41}
}

@ARTICLE{2014Gonzalez,
        author = {{Gonz\'{a}lez-Alfonso}, E. and {Fischer}, J. and {Grac\'{i}a-Carpio}, J. and {et al.}},
        title = "{The Mrk 231 molecular outflow as seen in OH}",
      journal = {\aap},
         year = 2014,
       volume = {561},
        pages = {A27}
}

@ARTICLE{2011Sturm,
        author = {{Sturm}, E. and {Gonz\'{a}lez-Alfonso}, E. and {Veilleux}, S. and {et al.}},
        title = "{Massive Molecular Outflows and Negative Feedback in ULIRGs Observed by Herschel/PACS}",
      journal = {\apjl},
         year = 2011,
         month = may,
       volume = {733},
        pages = {L16}
}

@ARTICLE{2011Matsuura,
        author = {{Matsuura}, M. and {Dwek}, E. and {Meixner}, M. and {et al.}},
        title = "{Herschel Detects a Massive Dust Reservoir in Supernova 1987A}",
      journal = {Science},
         year = 2011,
         month = sep,
       volume = {333},
        pages = {1258}
}

@ARTICLE{2015Matsuura,
        author = {{Matsuura}, M. and {Dwek}, E. and {Barlow}, M.~J. and {et al.}},
        title = "{A Stubbornly Large Mass of Cold Dust in the Ejecta of Supernova 1987A}",
      journal = {\apj},
         year = 2015,
         month = feb,
       volume = {800},
        pages = {50}
}

@ARTICLE{2013Barlow,
        author = {{Barlow}, M.~J. and {Swinyard}, B.~M. and {Owen}, P.~J. and {et al.}},
        title = "{Detection of a Noble Gas Molecular Ion, $^{36}$ArH$^+$, in the Crab Nebula}",
      journal = {Science},
         year = 2013,
         month = dec,
       volume = {342},
        pages = {1343}
}

@ARTICLE{2010Andre,
        author = {{Andr\'{e}}, Ph. and {Men'shchikov}, A. and {Bontemps}, S. and {et al.}},
        title = "{From filamentary clouds to prestellar cores to the stellar IMF: Initial highlights from the Herschel Gould Belt Survey}",
      journal = {\aap},
         year = 2010,
         month = jul,
       volume = {518},
        pages = {L102}
}

@ARTICLE{2010Menshchikov,
        author = {{Men'shchikov}, A. and {Andr\'{e}}, Ph. and {Didelon}, P. and {et al.}},
        title = "{Filamentary structures and compact objects in the Aquila and Polaris clouds observed by Herschel}",
      journal = {\aap},
         year = 2010,
         month = jul,
       volume = {518},
        pages = {L103}
}

@ARTICLE{2011Arzoumanian,
        author = {{Arzoumanian}, D. and {Andr\'{e}}, Ph. and {Didelon}, P. and {et al.}},
        title = "{Characterizing interstellar filaments with Herschel in IC 5146}",
      journal = {\aap},
         year = 2011,
       volume = {529},
        pages = {L6}
}

@ARTICLE{2019Arzoumanian,
        author = {{Arzoumanian}, D. and {Andr\'{e}}, Ph. and {K\"{o}nyves}, V. and {et al.}},
        title = "{Characterizing the properties of nearby molecular filaments observed with Herschel}",
      journal = {\aap},
         year = 2019,
       volume = {621},
        pages = {A42}
}

@ARTICLE{2010Konyves,
        author = {{K\"{o}nyves}, V. and {Andr\'{e}}, Ph. and {Men'shchikov}, A. and {et al.}},
        title = "{The Aquila prestellar core population revealed by Herschel}",
      journal = {\aap},
         year = 2010,
         month = jul,
       volume = {518},
        pages = {L106}
}

@ARTICLE{2015Konyves,
        author = {{K\"{o}nyves}, V. and {Andr\'{e}}, Ph. and {Men'shchikov}, A. and {et al.}},
        title = "{A census of dense cores in the Aquila cloud complex: SPIRE/PACS observations from the Herschel Gould Belt survey}",
      journal = {\aap},
         year = 2015,
       volume = {584},
        pages = {A91}
}

@ARTICLE{2020Konyves,
        author = {{K\"{o}nyves}, V. and {Andr\'{e}}, Ph. and {Arzoumanian}, D. and {et al.}},
        title = "{Properties of the dense core population in Orion B as seen by the Herschel Gould Belt survey}",
      journal = {\aap},
         year = 2020,
       volume = {635},
        pages = {A34}
}

@ARTICLE{2013Schneider,
        author = {{Schneider}, N. and {Andr\'{e}}, Ph. and {K\"{o}nyves}, V. and {et al.}},
        title = "{What Determines the Density Structure of Molecular Clouds? A Case Study of Orion B with Herschel}",
      journal = {\apjl},
         year = 2013,
         month = apr,
       volume = {766},
        pages = {L17}
}

@ARTICLE{2010WardThompson,
        author = {{Ward-Thompson}, D. and {Kirk}, J.~M. and {Andr\'{e}}, Ph. and {et al.}},
        title = "{A Herschel study of the properties of starless cores in the Polaris Flare dark cloud region using PACS and SPIRE}",
      journal = {\aap},
         year = 2010,
         month = jul,
       volume = {518},
        pages = {L102}
}

@ARTICLE{2013Konyves,
        author = {{K\"{o}nyves}, V. and {Andr\'{e}}, Ph. and {Schneider}, N. and {et al.}},
        title = "{Growing evidence for a core formation threshold traced in Herschel Gould Belt survey clouds}",
      journal = {Astron. Nachr.},
         year = 2013,
       volume = {334},
        pages = {908}
}

@INPROCEEDINGS{2015Andre,
       author = {{Andr\'{e}}, Ph.},
        title = "{The Herschel View of Star Formation}",
    booktitle = {XXVIIIth IAU General Assembly: Highlights of Astronomy, Vol. 16},
         year = 2015,
       editor = {{Montmerle}, T.},
       series = {University of Cambridge Press},
        pages = {31}       
}

@INPROCEEDINGS{2014Andre,
       author = {{Andr\'{e}}, Ph. and {Di Francesco}, J. and {Ward-Thompson}, D. and {et al.}},
        title = "{From Filamentary Networks to Dense Cores in Molecular Clouds: Toward a New Paradigm for Star Formation}",
    booktitle = {Protostars and Planets VI},
         year = 2014,
       editor = {{Beuther}, H. and {Klessen}, R.~S. and {Dullemond}, C.~P. and {Henning}, Th.},
       series = {The University of Arizona Press in collaboration with the Lunar and Planetary Institute},
        pages = {27}       
}

@ARTICLE{2010Motte,
        author = {{Motte}, F. and {Zavagno}, A. and {Bontemps}, S. and {et al.}},
        title = "{Initial highlights of the HOBYS key program, the Herschel imaging survey of OB young stellar objects}",
      journal = {\aap},
         year = 2010,
         month = jul,
       volume = {518},
        pages = {L77}
}

@ARTICLE{2010Zavagno,
        author = {{Zavagno}, A. and {Russeil}, D. and {Motte}, F. and {et al.}},
        title = "{Star formation triggered by the Galactic H II region RCW 120. First results from the Herschel Space Observatory}",
      journal = {\aap},
         year = 2010,
         month = jul,
       volume = {518},
        pages = {L81}
}

@ARTICLE{2010Bergin,
        author = {{Bergin}, E.~A. and {Phillips}, T.~G. and {Comito}, C. and {et al.}},
        title = "{Herschel observations of EXtra-Ordinary Sources (HEXOS): The present and future of spectral surveys with Herschel/HIFI}",
      journal = {\aap},
         year = 2010,
         month = oct,
       volume = {521},
        pages = {L20}
}

@ARTICLE{2014Crockett,
        author = {{Crockett}, N.~R. and {Bergin}, E.~A. and {Neill}, J.~L. and {et al.}},
        title = "{Herschel Observations of EXtraOrdinary Sources: Analysis of the HIFI 1.2 THz Wide Spectral Survey Toward Orion KL. I. Methods}",
      journal = {\apj},
         year = 2014,
         month = jun,
       volume = {787},
        pages = {112}
}

@ARTICLE{2015Crockett,
        author = {{Crockett}, N.~R. and {Bergin}, E.~A. and {Neill}, J.~L. and {et al.}},
        title = "{Herschel Observations of EXtraOrdinary Sources: Analysis of the HIFI 1.2 THz Wide Spectral Survey Toward Orion KL. II. Chemical Implications}",
      journal = {\apj},
         year = 2015,
         month = jun,
       volume = {806},
        pages = {239}
}

@ARTICLE{2010aMolinari,
        author = {{Molinari}, S. and {Swinyard}, B. and {Bally}, J. and {et al.}},
        title = "{Hi-GAL: The Herschel Infrared Galactic Plane Survey}",
      journal = {\pasp},
         year = 2010,
         month = mar,
       volume = {122},
        pages = {314}
}

@ARTICLE{2010bMolinari,
        author = {{Molinari}, S. and {Swinyard}, B. and {Bally}, J. and {et al.}},
        title = "{Clouds, filaments, and protostars: The Herschel Hi-GAL Milky Way}",
      journal = {\aap},
         year = 2010,
         month = jul,
       volume = {518},
        pages = {L100}
}

@INPROCEEDINGS{2014Molinari,
       author = {{Molinari}, S. and {Bally}, J. and {Glover}, S. and {et al.}},
        title = "{The Milky Way as a Star Formation Engine}",
    booktitle = {Protostars and Planets VI},
         year = 2014,
       editor = {{Beuther}, H. and {Klessen}, R.~S. and {Dullemond}, C.~P. and {Henning}, Th.},
       series = {The University of Arizona Press in collaboration with the Lunar and Planetary Institute},
        pages = {125}       
}

@ARTICLE{2016aMolinari,
        author = {{Molinari}, S. and {Schisano}, E. and {Elia}, D. and {et al.}},
        title = "{Hi-GAL, the Herschel infrared Galactic Plane Survey: photometric maps and compact source catalogues}",
      journal = {\aap},
         year = 2016,
       volume = {591},
        pages = {A149}
}

@ARTICLE{2017Elia,
        author = {{Elia}, D. and {Molinari}, S. and {Schisano}, E. and {et al.}},
        title = "{The Hi-GAL compact source catalogue – I. The physical properties of the clumps in the inner Galaxy (−71$^{\circ}$.0 < l < 67$^{\circ}$.0)}",
      journal = {\mnras},
         year = 2017,
         month = jun,
       volume = {471},
        pages = {100}
}

@ARTICLE{2021Elia,
        author = {{Elia}, D. and {Merello}, M. and {Molinari}, S. and {et al.}},
        title = "{The Hi-GAL compact source catalogue – II. The 360$^{\circ}$ catalogue of clump physical properties}",
      journal = {\mnras},
         year = 2021,
         month = apr,
       volume = {504},
        pages = {2742}
}

@ARTICLE{2016bMolinari,
        author = {{Molinari}, S. and {Noriega-Crespo}, A. and {Bally}, J. and {et al.}},
        title = "{Large-scale latitude distortions of the inner Milky Way disk from the Herschel/Hi-GAL Survey}",
      journal = {\aap},
         year = 2016,
       volume = {588},
        pages = {A75}
}

@ARTICLE{2018Elia,
        author = {{Elia}, D. and {Strafella}, F. and {Dib}, S. and {et al.}},
        title = "{Multifractal analysis of the interstellar medium: first application to Hi-GAL observations}",
      journal = {\mnras},
         year = 2018,
         month = aug,
       volume = {481},
        pages = {509}
}

@ARTICLE{2010Langer,
        author = {{Langer}, W.~D. and {Velusamy}, T. and {Pineda}, J.~L. and {et al.}},
        title = "{C$^+$ detection of warm dark gas in diffuse clouds}",
      journal = {\aap},
         year = 2010,
       volume = {521},
        pages = {L17}
}

@ARTICLE{2013Pineda,
        author = {{Pineda}, J.~L. and {Langer}, W.~D. and {Velusamy}, T. and {et al.}},
        title = "{A Herschel [C II] Galactic plane survey I. The global distribution of ISM gas components}",
      journal = {\aap},
         year = 2013,
       volume = {554},
        pages = {A103}
}

@ARTICLE{2014Langer,
        author = {{Langer}, W.~D. and {Velusamy}, T. and {Pineda}, J.~L. and {et al.}},
        title = "{A Herschel [C II] Galactic plane survey II. CO-dark H$_2$ in clouds}",
      journal = {\aap},
         year = 2014,
       volume = {561},
        pages = {A122}
}

@ARTICLE{2014Pineda,
        author = {{Pineda}, J.~L. and {Langer}, W.~D. and {Goldsmith}, P.~F. and {et al.}},
        title = "{A Herschel [C II] Galactic plane survey III. [C II] as a tracer of star formation}",
      journal = {\aap},
         year = 2014,
       volume = {570},
        pages = {A121}
}

@ARTICLE{2010Eiroa,
        author = {{Eiroa}, C. and {Fedele}, D. and {Maldonado}, J. and {et al.}},
        title = "{Cold DUst around NEarby Stars (DUNES). First results. A resolved exo-Kuiper belt around the solar-like star ${\zeta}^2$ Ret}",
      journal = {\aap},
         year = 2010,
         month = jul,
       volume = {518},
        pages = {L131}
}

@ARTICLE{2010Matthews,
        author = {{Matthews}, B.~C. and {Sibthorpe}, B. and {Kennedy}, G. and {et al.}},
        title = "{Resolving debris discs in the far-infrared: Early highlights from the DEBRIS survey}",
      journal = {\aap},
         year = 2010,
         month = jul,
       volume = {518},
        pages = {L135}
}

@ARTICLE{2010Sibthorpe,
        author = {{Sibthorpe}, B. and {Vandenbussche}, B. and {Greaves}, J.~S. and {et al.}},
        title = "{The Vega debris disc: A view from Herschel}",
      journal = {\aap},
         year = 2010,
         month = jul,
       volume = {518},
        pages = {L130}
}

@ARTICLE{2010Liseau,
        author = {{Liseau}, R. and {Eiroa}, C. and {Fedele}, D. and {et al.}},
        title = "{Resolving the cold debris disc around a planet-hosting star. PACS photometric imaging observations of q$^1$ Eridani (HD 10647, HR 506)}",
      journal = {\aap},
         year = 2010,
         month = jul,
       volume = {518},
        pages = {L132}
}

@ARTICLE{2010Vandenbussche,
        author = {{Vandenbussche}, B. and {Sibthorpe}, B. and {Acke}, B. and {et al.}},
        title = "{The $\beta$ Pictoris disk imaged by Herschel PACS and SPIRE}",
      journal = {\aap},
         year = 2010,
         month = jul,
       volume = {518},
        pages = {L133}
}

@ARTICLE{2013Eiroa,
        author = {{Eiroa}, C. and {Marshall}, J.~P. and {Mora}, A. and {et al.}},
        title = "{DUst around NEarby Stars. The survey observational results}",
      journal = {\aap},
         year = 2013,
       volume = {555},
        pages = {A11}
}

@ARTICLE{2016Montesinos,
        author = {{Montesinos}, B. and {Eiroa}, C. and {Krivov}, A.~V. and {et al.}},
        title = "{Incidence of debris discs around FGK stars in the solar neighbourhood}",
      journal = {\aap},
         year = 2016,
       volume = {593},
        pages = {A51}
}

@ARTICLE{2012Acke,
        author = {{Acke}, B. and {Min}, M. and {Dominik}, C. and {et al.}},
        title = "{Herschel images of Fomalhaut. An extrasolar Kuiper belt at the height of its dynamical activity}",
      journal = {\aap},
         year = 2012,
       volume = {540},
        pages = {A125}
}

@ARTICLE{2012deVries,
        author = {{de Vries}, B.~L. and {Acke}, B. and {Blommaert}, A.~D.~L. and {et al.}},
        title = "{Comet-like mineralogy of olivine crystals in an extrasolar proto-Kuiper belt}",
      journal = {\nat},
         year = 2012,
         month = oct,
       volume = {490},
        pages = {74}
}

@ARTICLE{2012aDecin,
        author = {{Decin}, L. and {Cox}, N.~L.~J. and {Royer}, P. and {et al.}},
        title = "{The enigmatic nature of the circumstellar envelope and bow shock surrounding Betelgeuse as revealed by Herschel. I. Evidence of clumps, multiple arcs, and a linear bar-like structure}",
      journal = {\aap},
         year = 2012,
       volume = {548},
        pages = {A113}
}

@ARTICLE{2010Decin,
        author = {{Decin}, L. and {Ag\'{u}ndez}, M. and {Barlow}, M.~J. and {et al.}},
        title = "{Warm water vapour in the sooty outflow from a luminous carbon star}",
      journal = {\nat},
         year = 2010,
         month = sep,
       volume = {467},
        pages = {64}
}

@ARTICLE{2012bDecin,
        author = {{Decin}, L.},
        title = "{Late stages of stellar evolution – Herschel’s contributions}",
      journal = {Adv. Space Res.},
         year = 2012,
        month = jun,
       volume = {50},
        pages = {843}
}

@INPROCEEDINGS{2014vanDishoeck,
       author = {{van Dishoeck}, E.~F. and {Bergin}, E.~A. and {Lis}, D.~C. and {Lunine}, J.~I.},
        title = "{Water: From Clouds to Planets}",
    booktitle = {Protostars and Planets VI},
         year = 2014,
       editor = {{Beuther}, H. and {Klessen}, R.~S. and {Dullemond}, C.~P. and {Henning}, Th.},
       series = {The University of Arizona Press in collaboration with the Lunar and Planetary Institute},
        pages = {835}       
}

@ARTICLE{2012Caselli,
        author = {{Caselli}, P. and {Keto}, E. and {Bergin}, E.~A. and {et al.}},
        title = "{First Detection of Water Vapour in a Pre-stellar Core}",
      journal = {\apjl},
         year = 2012,
        month = nov,
       volume = {759},
        pages = {L37}
}

@ARTICLE{2012Kristensen,
        author = {{Kristensen}, L.~E. and {van Dishoeck}, E.~F. and {Bergin}, E.~A. and {et al.}},
        title = "{Water in star-forming regions with Herschel (WISH). II. Evolution of 557 GHz 1$_{10}$–1$_{01}$ emission in low-mass protostars}",
      journal = {\aap},
         year = 2012,
       volume = {542},
        pages = {A8}
}

@ARTICLE{2011Hogerheijde,
        author = {{Hogerheijde}, M.~R. and {Bergin}, E.~A. and {Brinch}, C. and {et al.}},
        title = "{Detection of the Water Reservoir in a Forming Planetary System}",
      journal = {Science},
         year = 2011,
         month = oct,
       volume = {334},
        pages = {338}
}

@ARTICLE{2013Bergin,
        author = {{Bergin}, E.~A. and {Cleeves}, L.~I. and {Gorti}, U. and {et al.}},
        title = "{An old disk still capable of forming a planetary system}",
      journal = {\nat},
         year = 2013,
         month = jan,
       volume = {493},
        pages = {644}
}

@ARTICLE{2013Cavalie,
        author = {{Cavali\'{e}}, T. and {Feuchtgruber}, H. and {Lellouch}, E. and {et al.}},
        title = "{Spatial distribution of water in the stratosphere of Jupiter from Herschel HIFI and PACS observations}",
      journal = {\aap},
         year = 2013,
       volume = {553},
        pages = {A21}
}

@ARTICLE{2011aHartogh,
        author = {{Hartogh}, P. and {Lellouch}, E. and {Moreno}, R. and {et al.}},
        title = "{Direct detection of the Enceladus water torus with Herschel}",
      journal = {\aap},
         year = 2011,
       volume = {532},
        pages = {L2}
}

@ARTICLE{2019Cavalie,
        author = {{Cavali\'{e}}, T. and {Hue}, V. and {Hartogh}, P. and {et al.}},
        title = "{Herschel map of Saturn’s stratospheric water, delivered by the plumes of Enceladus}",
      journal = {\aap},
         year = 2019,
       volume = {630},
        pages = {A87}
}

@ARTICLE{2014Kuppers,
        author = {{K\"{u}ppers}, M. and {O'Rourke}, L. and {Bockel\'{e}e-Morvan}, D. and {et al.}},
        title = "{Localized sources of water vapour on the dwarf planet (1) Ceres}",
      journal = {\nat},
         year = 2014,
         month = jan,
       volume = {505},
        pages = {525}
}

@ARTICLE{2010Hartogh,
        author = {{Hartogh}, P. and {Crovisier}, J. and {de Val-Borro}, M. and {et al.}},
        title = "{HIFI observations of water in the atmosphere of comet C/2008 Q3 (Garradd)}",
      journal = {\aap},
         year = 2010,
         month = jul,
       volume = {518},
        pages = {L150}
}

@ARTICLE{2011bHartogh,
        author = {{Hartogh}, P. and {Lis}, D. and {Bockel\'{e}e-Morvan}, D. and {et al.}},
        title = "{Ocean-like water in the Jupiter-family comet 103P/Hartley 2}",
      journal = {\nat},
         year = 2011,
         month = oct,
       volume = {478},
        pages = {218}
}

@ARTICLE{2012BockeleeMorvan,
        author = {{Bockel\'{e}e-Morvan}, D. and {Biver}, N. and {Swinyard}, B. and {et al.}},
        title = "{Herschel measurements of the D/H and $^{16}$O/$^{18}$O ratios in water in the Oort-cloud comet C/2009 P1 (Garradd)}",
      journal = {\aap},
         year = 2012,
       volume = {544},
        pages = {L15}
}

@ARTICLE{2013Lis,
        author = {{Lis}, D.~C. and {Biver}, N. and {Bockel\'{e}e-Morvan}, D. and {et al.}},
        title = "{A Herschel Study of D/H in Water in the Jupiter-family Comet 45P/Honda-Mrkos-Pajdu\v{s}\'{a}kov\'{a} and Prospects for D/H Measurements with CCAT}",
      journal = {\apjl},
         year = 2013,
         month = sep,
       volume = {774},
        pages = {L3}
}

@ARTICLE{2015Altwegg,
        author = {{Altwegg}, K. and {Balsiger}, H. and {Bar-Nun}, A. and {et al.}},
        title = "{67P/Churyumov-Gerasimenko, a Jupiter family comet with a high D/H ratio}",
      journal = {Science},
         year = 2015,
         month = jan,
       volume = {347},
        pages = {1261952}
}

@ARTICLE{2012Vilenius,
        author = {{Vilenius}, E. and {Kiss}, C. and {Mommert}, M. and {et al.}},
        title = "{“TNOs are Cool”: A survey of the trans-Neptunian region. VI. Herschel/PACS observations and thermal modeling of 19 classical Kuiper belt objects}",
      journal = {\aap},
         year = 2012,
       volume = {541},
        pages = {A94}
}

@ARTICLE{2014Vilenius,
        author = {{Vilenius}, E. and {Kiss}, C. and {M\"{u}ller}, T. and {et al.}},
        title = "{“TNOs are Cool”: A survey of the trans-Neptunian region. X. Analysis of classical Kuiper belt objects from Herschel and Spitzer observations}",
      journal = {\aap},
         year = 2014,
       volume = {564},
        pages = {A35}
}

@ARTICLE{2014Lacerda,
        author = {{Lacerda}, P. and {Fornasier}, S. and {Lellouch}, E. and {et al.}},
        title = "{The Albedo-Color Diversity of Transneptunian Objects}",
      journal = {\apjl},
         year = 2014,
       volume = {793},
        pages = {L2}
}

@ARTICLE{2016Lellouch,
        author = {{Lellouch}, E. and {Santos-Sanz}, P. and {Fornasier}, S. and {et al.}},
        title = "{The long-wavelength thermal emission of the Pluto-Charon system from Herschel observations. Evidence for emissivity effects}",
      journal = {\aap},
         year = 2016,
       volume = {588},
        pages = {A2}
}

@ARTICLE{2021Shirley,
        author = {{Shirley}, R. and {Duncan}, K. and {Campos Varillas}, M.~C. and {et al.}},
        title = "{HELP: the Herschel Extragalactic Legacy Project}",
      journal = {\mnras},
         year = 2021,
       volume = {507},
        pages = {129}
}

@ARTICLE{2019Sciacca,
        author = {{Sciacca}, E. and {Vitello}, D. and {Becciani}, U. and {et al.}},
        title = "{VIALACTEA science gateway for Milky Way analysis}",
      journal = {Future Generation Computer Systems},
         year = 2021,
       volume = {94},
        pages = {947}
}

@ARTICLE{2017Davies,
        author = {{Davies}, J.~I. and {Baes}, M. and {Bianchi}, S. and {et al.}},
        title = "{DustPedia: A Definitive Study of Cosmic Dust in the Local Universe}",
      journal = {\pasp},
         year = 2017,
       volume = {129},
        pages = {044102}
}

@ARTICLE{2018Muller,
        author = {{M\"{u}ller}, T.~G. and {Marciniak}, A. and {Kiss}, Cs. and {et al.}},
        title = "{Small Bodies Near and Far (SBNAF): A benchmark study on physical and thermal properties of small bodies in the Solar System}",
      journal = {Adv. Space Res.},
         year = 2018,
       volume = {62},
        pages = {2326}
}

@ARTICLE{1995Fleck,
       author = {{Fleck}, B. and {Domingo}, V. and {Poland}, A.~I.},
        title = "{The SOHO mission}",
      journal = {Solar Physics},
         year = 1995,
        month = dec,
       volume = {162},
       number = {1}
}

@ARTICLE{2010Tauber,
       author = {{Tauber}, J.~A. and {Mandolesi}, N. and {Puget}, J. -L. and {Banos}, T. and {Bersanelli}, M. and {Bouchet}, F.~R. and {Butler}, R.~C. and {Charra}, J. and {Crone}, G. and {Dodsworth}, J. and {Efstathiou}, G. and {Gispert}, R. and {Guyot}, G. and {Gregorio}.},
        title = "{Planck pre-launch status: The Planck mission}",
      journal = {\aap},
         year = 2010,
        month = sep,
       volume = {520},
        pages = {A1}
}

@ARTICLE{2001Escoubet,
       author = {{Escoubet}, C.~P. and {Fehringer}, M. and {Goldstein}, M.},
        title = "{Introduction: The Cluster mission}",
      journal = {Annales Geophysicae},
         year = 2001,
        month = oct,
       volume = {19},
        pages = {1197-1200}
}

@article{2009Glassmeier,
  title={Rosetta -- ESA's mission to the origin of the solar system},
  author={{Glassmeier}, K.-H. and {Boehnhardt}, H. and {Koschny}, D. and {K{\"u}hrt}, E. and {Richter}, I.},
  year={2009}
}

@ARTICLE{2024XMM,
       author = {{Parmar}, A.N. and {Schartel}, N. and {Santos-Lleó} M.},
        title = "ESA Science Programme Missions: Contributions and Exploitation -- XMM-Newton Observing Time Proposals",
      journal = {Submitted to ArXiv},
         year = 2024,
         volume = {},
         page={}
}

@ARTICLE{2024INTEGRAL,
       author = {{Kuulkers}, E. and {Sánchez-Fernández}, C. and {Parmar}, A.N.},
        title = "ESA Science Programme Missions: Contributions and Exploitation -- INTEGRAL Observing Time Proposals",
      journal = {Submitted to ArXiv},
         year = 2024,
         volume = {},
         page={}
}

@ARTICLE{2024pubs,
       author = {{De Marchi}, G. and {Parmar}, A.N.},
        title = "ESA Science Programme Missions: Contributions and Exploitation -- ESA Mission Publications",
      journal = {Submitted to ArXiv},
         year = 2024,
         volume = {},
         page={}
}

\clearpage

\section*{\large{Acronym List}}
\addcontentsline{toc}{section}{\large{Acronym List}}

\begin{acronym}[MPCCCCCC] 

\acro{AAS}{American Astronomical Society}
\acro{ADS}{Astrophysics Data Service}
\acro{AO}{Announcement of Opportunity}
\acro{ALMA}{Atacama Large Millimeter/sub-millimeter Array}
\acro{CSG}{Centre Spatial Guyanais}
\acro{D-SCI}{ESA Director of Science}
\acro{DDT}{Director's Discretionary Time}
\acro{ESA}{European Space Agency}
\acro{ESAC}{European Space Astronomy Centre}
\acro{ESOC}{European Space Operations Centre}
\acro{FIRST}{Far Infra-Red and Submillimetre space Telescope}
\acro{FOV}{Field of View}
\acro{GT}{Guaranteed Time}
\acro{G/AGN}{galaxies and active galactic nuclei}
\acro{HCSS}{Herschel Common Science System}
\acro{HIFI}{Heterodyne Instrument for the Far Infrared}
\acro{HIPE}{Herschel Interactive Processing Environment}
\acro{HOTAC}{Herschel Observing Time Allocation Committee}
\acro{HSA}{Herschel Science Archive}
\acro{HSC}{Herschel Science Centre}
\acro{HST}{Hubble Space Telescope}
\acro{HUG}{Herschel Users' Group}
\acro{IAU}{International Astronomical Union}
\acro{ICC}{Instrument Control Centre}
\acro{IPAC}{Infrared Processing and Analysis Center}
\acro{ISM}{Interstellar Medium}
\acro{ISM/SF/SS}{Interstellar Medium, Star Formation and Solar System Objects}
\acro{ISO}{Infrared Space Observatory}
\acro{ISOCAM}{ISO Infrared Camera}
\acro{ISOPHOT}{ISO Photo-polarimeter}
\acro{JWST}{James Webb Space Telescope}
\acro{KP}{Key Programme}
\acro{KPI}{Key Performance Indicator}
\acro{KPs}{Key Programmes}
\acro{KPGT}{Key Programme Guaranteed Time}
\acro{KPOT}{Key Programme Open Time}
\acro{LWS}{Long Wave Spectrometer}
\acro{MOC}{Mission Operations Centre}
\acro{MPE}{Max Planck Institute for Extraterrestrial Physics}
\acro{NASA}{National Aeronautics and Space Administration}
\acro{NHSC}{NASA Herschel Science Center}
\acro{NUP}{NHSC Users' Panel}
\acro{OT}{Open Time}
\acro{PACS}{Photodetector Array Camera and Spectrometer}
\acro{PhD}{Doctor of Philosophy Degree}
\acro{PI}{Principal Investigator}
\acro{PS}{Project Scientist}
\acro{RAL}{Rutherford Appleton Laboratory}
\acro{SDP}{Science Demonstration Phase}
\acro{SGS}{Science Ground Segment}
\acro{SMP}{Science Management Plan}
\acro{SOC}{Science Operations Centre}
\acro{SPC}{Science Programme Committee}
\acro{SPIRE}{Spectral and Photometric Imaging REciever}
\acro{SRON}{Netherlands Institute for Space Research}
\acro{S/SE}{Stars and Stellar Evolution}
\acro{SWAS}{Submillimeter Wave Astronomy Satellite}
\acro{SWS}{Short Wave Spectrometer}
\acro{TNO}{Trans-Neptunian Object}
\acro{ULIRGs}{Ultra Luminous InfraRed Galaxies}

\end{acronym}

\end{document}